\documentclass[aps,prd,reprint,twocolumn,preprintnumbers,floatfix,nofootinbib]{revtex4-1}

\pdfoutput=1

\usepackage{ragged2e}
\usepackage{subfig}
\usepackage{graphicx}
\usepackage{epsfig}
\usepackage{amsmath}
\usepackage{empheq}
\usepackage{amsfonts}
\usepackage{amssymb}
\usepackage{color}
\usepackage[bottom]{footmisc}
\usepackage{array,multirow,makecell}
\usepackage{slashed}
\usepackage[colorlinks,citecolor=blue]{hyperref}
\usepackage{float}
\usepackage{ulem}
\usepackage[utf8]{inputenc}
\usepackage[english]{babel}
\usepackage{url}
\usepackage{hyperref}
\usepackage{xcolor}
\usepackage{lipsum}
\usepackage[bottom]{footmisc}
\usepackage{pifont}

\usepackage[most]{tcolorbox}
\newtcbox{\mymath}[1][]{%
    nobeforeafter, math upper, tcbox raise base,
    enhanced, colframe=blue!30!black,
    colback=blue!30, boxrule=1pt,
    #1}

\usepackage{blindtext}

\def\lsim{\raise0.3ex\hbox{$\;<$\kern-0.75em\raise-1.1ex\hbox{$\sim\;$}}}
\def\gsim{\raise0.3ex\hbox{$\;>$\kern-0.75em\raise-1.1ex\hbox{$\sim\;$}}}
\def\nn{\nonumber}

\newcommand{\bea}{\begin{aligned}}
\newcommand{\eea}{\end{aligned}}
\def\beq{\begin{equation}}
\def\eeq{\end{equation}}
\def\beqa{\begin{eqnarray}}
\def\eeqa{\end{eqnarray}}
\def\be{\begin{equation}}
\def\ee{\end{equation}}

\def\bse{\begin{subequations}}
\def\ese{\end{subequations}}

\def\bea{\begin{eqnarray}}
\def\eea{\end{eqnarray}}

\newcommand{\baz}{\begin{array}{cc}}

\newcommand{\bav}{\begin{array}{cccc}}

\newcommand{\noi}{\noindent}

\begin{document}

\title{Gravitational ultra-relativistic freeze-out during general reheating}

\author{Mathieu Gross}
\email{mathieu.gross@ijclab.in2p3.fr}
\affiliation{Universit\'e Paris-Saclay, CNRS/IN2P3, IJCLab, 91405 Orsay, France}

\author{Stephen E. Henrich}
    \email{henri455@umn.edu}
    \affiliation{William I. Fine Theoretical Physics Institute, School of Physics and Astronomy, University of Minnesota, Minneapolis, Minnesota 55455, USA}
    
\author{Fotis Koutroulis}
\email{fkoutroulis@ihep.ac.cn}
\affiliation{Theoretical Physics Division, Institute of High Energy Physics, Chinese Academy of Sciences,
19B Yuquan Road, Shijingshan District, Beijing 100049, China}
\affiliation{China Center of Advanced Science and Technology, Beijing 100190, China}

\begin{abstract}

We investigate ultrarelativistic freeze-out (UFO) in the context of generic reheating scenarios. 
While the standard WIMP dark matter paradigm has been extensively studied, UFO has 
so far only been analyzed within the specific reheating channel $\phi \rightarrow f\bar{f}$. 
Unlike in the standard WIMP mechanism, where dark matter can only be diluted  after freeze-out at 
$T_\mathrm{FO} \sim m_\chi/\mathcal{O}(10)$, UFO dark matter can undergo freeze-in like 
phases following the initial freeze-out, driven by the non-trivial temperature evolution. The exact temperature evolution then needs to be accounted for, as a change in the temperature scaling can modify the IR/UV nature of UFO, impacting the relic abundance. We first generalize UFO  to an arbitrary temperature profile 
$T \sim a^{-\xi}$, making explicit the UV and IR regimes for a thermally averaged 
cross section $\langle \sigma v \rangle \sim T^n / \Lambda^{n+2}$. Then, as a concrete example, we consider the minimal scenario 
in which gravitational particle production at the onset of reheating sources an initial radiation 
abundance, and show that this early hot bath changes the UFO parameter space. We refer to this effect as GUFO. Specializing to $n = 2$, we find that matter-like reheating ($V \sim \phi^2$) 
accommodates dark matter masses up to $10^7~\mathrm{GeV}$ for 
$\Lambda \lesssim 10^9~\mathrm{GeV}$ as thermalization becomes less stringent, while radiation-like 
reheating ($V \sim \phi^4$) is compatible with GUFO across all reheating channels only if gravitational processes are taken into account. 

\end{abstract}

\maketitle

Dark matter is one of the cornerstones of modern cosmology, as it is required to explain several observables, such as structure formation and the cosmic microwave background. The main effort now concerns both its nature, either as a particle \cite{Cirelli:2024ssz,Arcadi:2024ukq,Bernal:2017kxu} or wave-like condensation within galactic scales \cite{Ferreira:2020fam} or as primordial black holes \cite{Carr:2021bzv}, and its proper production mechanism. From a particle physics perspective, significant theoretical and experimental efforts in recent years have been devoted to understand and constrain possible production scenarios. The main paradigms are then separated depending on the strength of the interaction. For strongly coupled scenarios, the particle is believed to reach thermal equilibrium in the early universe, leading to the standard WIMP scenario \cite{Arcadi:2024ukq}. For feebly interacting models (small couplings and/or large mediators), the dark matter remains out-of-equilibrium and its abundance gradually accumulates over time to reach the measured relic abundance today $\Omega_{\chi}h^{2}\simeq 0.12$ \cite{Bernal:2017kxu,Planck:2018vyg}. This mechanism is called freeze-in, and the associated particles are commonly called FIMPs. These paradigms have driven experimental searches for many years and continue to do so today \cite{LZ:2022lsv,XENON:2024znc,XENON:2023cxc,DARWIN:2016hyl,McDaniel:2023bju,Gaskins:2016cha}. However, the lack of experimental evidence now motivates a revisit and extension of these scenarios to other cosmological contexts such as pure gravitational interaction \cite{Clery:2021bwz,Clery:2022wib,Barman:2022qgt,Kolb:2025wyj,Jenks:2024fiu,Kolb:2023ydq, Koutroulis:2023fgp, Feiteira:2025phi,Verner:2025ise} or reheating-era production \cite{Bernal:2022wck,Belanger:2026ctm,Henrich:2025gsd,Henrich:2025sli,Henrich:2025pca,Cosme:2023xpa,Feiteira:2026qme,Arcadi:2024wwg,Henrich:2024rux}. In the latter context a new way to produce dark matter through ultra relativistic freeze-out (UFO) was noticed \cite{Henrich:2025gsd,Henrich:2025sli,Henrich:2025pca}. This scenario, first introduced with neutrinos, is normally forbidden in radiation domination due to structure formation constraints. During reheating, however, it becomes allowed as the continuous entropy injection dilutes and cools the otherwise warm and overproduced dark matter. In this work, we generalize the UFO mechanism to generic reheating scenarios and we investigate the effects of gravitationally produced radiation.

 In addition to evading warmness constraints, UFO during reheating exhibits several other attractive features. First, the fact that UFO dark matter equilibrates erases any initial abundance that may have been produced gravitationally during inflation or immediately after. This relaxes the constraints on the reheating temperature ($T_{\rm{RH}}$) that would otherwise apply to freeze-in dark matter. Second, after (ultra)relativistic decoupling but prior to the completion of reheating, there may be significant out-of-equilibrium freeze-in like production as well as entropy dilution, which together open a vast parameter space in ($m_{\rm DM},T_{\rm RH},\Lambda$), with $\Lambda$ parametrically related to the mass of a heavy mediator
in the UV theory, compared to conventional freeze-in and freeze-out. Third, because UFO dark matter is allowed to thermalize, this mechanism is generally compatible with stronger interactions (lower $\Lambda$ in effective field theory) compared to standard freeze-in, which renders UFO more accessible to detection efforts than FIMPs \cite{Henrich:2026tox}. Finally, while UFO has the advantage of equilibrating, similar to WIMPs, it also exhibits UV or IR sensitivity depending on the specifics of the interaction and the cosmological history. This endows UFO with rich phenomenology in a similar manner to UV and IR freeze-in.

 We start by considering a generic beyond the Standard Model (BSM) interaction between dark matter particles $\chi$ and Standard Model particles involving the exchange of a heavy mediator of mass $M$. We typically have in mind $2\rightarrow 2$ annihilations between Standard Model fermions (or the Higgs) and dark matter. In the relativistic regime where $T \gg m_\chi,m_f$, the thermally averaged cross section  can be parametrized as
\begin{equation}
\label{eq:generic sigmav}
    \langle\sigma v\rangle\sim\frac{g^{4}\mu^{2-n}T^{n}}{(T^{2}+M^{2})^{2}}
\end{equation}
where $\mu$ is a dimensionful coupling, $g$ is the vertex coupling and $T$ is the temperature of the thermal bath. This cross section will then change its temperature-dependence based upon whether the temperature is higher or lower than the mediator mass. In each case, the expression reduces to $\langle\sigma v\rangle\sim T^{n}/\Lambda^{n+2}$ where the $n$ parameter will change between the high  ($n_{h}$) and low ($n_{l}$) energy regimes, which are linked by the relation $n_l = n_h+4$. This change of behavior will then allow a relativistic candidate to enter thermal equilibrium while $n=n_{h}$ and later on to exit equilibrium when $n=n_{l}$. We will now put this subtlety aside and refer to the low energy index as $n$. We note that the parametrization in Eq.~(\ref{eq:generic sigmav}) breaks down when the Standard Model bath temperature drops to approximately max($m_\chi,m_f$). At these low temperatures, there will typically be terms in the thermally averaged cross section proportional to powers of $m_\chi$ and $m_f$, such that the relativistic parametrization above is no longer appropriate. This is important when determining the boundary between relativistic and non-relativistic freeze-out, which has been explored in detail in \cite{Henrich:2025pca,Henrich:2026tox}. For this paper, we focus on dark matter decoupling and production in the relativistic regime.

In effective field theory, there is a straightforward relationship between the operator dimension and $n$. Specifically, for an effective $d$-dimensional operator $\mathcal{O}_{d}$ with $\mathcal{L}_{\rm eff}\supset \frac{\mathcal{O}_{d}}{\Lambda^{d-4}}$, the thermally averaged cross section in the relativistic regime scales as $\langle \sigma v \rangle \propto \frac{T^{2d-10}}{\Lambda^{2d-8}}$. This implies that $n=2d-10$. For instance, an effective 4-Fermi interaction with operator dimension 6 corresponds to $n=2$. For standard freeze-in during radiation domination, UV vs. IR-dominated production is distinguished by a critical value of $n_c=-1$. Thus, renormalizable operators generally correspond to IR freeze-in (sensitivity to the low temperature regime $T\approx m_\chi$) while non-renormalizable operators correspond to UV freeze-in (sensitivity to $T_{\rm RH}$). However, when reheating is treated carefully as a non-instantaneous process, the critical value of $n$ changes and the maximum temperature reached by the thermal bath is $T_{\rm max} \gg T_{\rm RH}$, such that not all non-renormalizable operators are sensitive to the highest temperatures. This will become important for our results below.

To be consistent with Ref.~\cite{Henrich:2025gsd}, we take the equilibration condition to be $\Gamma > t^{-1}$ with $t$ the cosmic time. This condition is rewritten in terms of the Hubble parameter as
\begin{equation}
\label{eq:thermal eq}
    \Gamma>\frac{3(1+w)}{2}H\,,
\end{equation}
for a generic $w$ domination era where $w$ is the equation of state of the species dominating the energy budget of the universe.

The paper is organized as follows. After reviewing the necessary reheating elements in Sec.~\ref{Sec:reheating}, we review the UFO phenomenology in a generic temperature background $T\sim a^{-\xi}$ focusing on the decoupling in Sec.~\ref{sec:UV_IR_UFO}. We discuss the impact of a potentially mixed temperature profile, taking as an example the gravitationally produced radiation in Sec.~\ref{sec:GUFO}. Finally we compute the relic abundance of dark matter including this effect in Sec.~\ref{sec:relic abundance} and adapt the necessary warm dark matter constraints in Sec.~\ref{sec:compatibilite LCDM} before concluding.

\section{Generic reheating}
\label{Sec:reheating}

During reheating the inflaton potential can be taylor expanded in a monomial form 
\begin{equation}
\label{eq:inflaton potential}
    V(\phi)\simeq \lambda M_{P}^{4}\left(\frac{\phi}{M_{P}}\right)^{k}\,,
\end{equation}
where $M_{P}=2.435\times 10^{18}$~GeV is the reduced Planck mass.

Our work is not sensitive to the details of the inflationary era but this expansion can for example be achieved considering the T-$\alpha$ attractor class of model~\cite{Kallosh:2013hoa}. In the perturbative approach, the inflaton field then acquires an average equation of state $w_{\phi} =(k-2)/(k+2)$ through the oscillations in this potential. The coupling of these oscillations to daughter fields sources the radiation bath whose domination is necessary for a consistent embedding of inflation \cite{Planck:2018vyg}. The oscillation-averaged system then evolves under the following set of coupled continuity equations
\begin{align}
    \dot{\rho}_{\phi}+3H(1+w_{\phi})\rho_{\phi} &\simeq -(1+w_{\phi})\Gamma_{\phi}\rho_{\phi}\,,\\
    \dot{\rho}_{R}+4H\rho_{R} &\simeq(1+w_{\phi})\Gamma_{\phi}\rho_{\phi}\,,
\end{align}
where $\Gamma_{\phi}$ is the inflaton decay rate stemming from the inflaton coupling.  Throughout this work, we will consider 3 different channels for reheating, namely  \footnote{Note that the $\mu
$ in this interaction differs from the one in the thermal-averaged cross section. Since neither is explicitly used in our results, we do not emphasize their difference.}
\begin{equation}
\label{eq:inflaton channels}
    \mathcal{L} \supset
\begin{cases}
y\,\phi\,\bar{f} f      & \phi \rightarrow \bar{f} f \\
\mu\,\phi\, b b         & \phi \rightarrow b b \\
\sigma\,\phi^{2} b^{2}  & \phi\phi \rightarrow b b
\end{cases}\,,
\end{equation}
with $f$ and $b$ fermionic and bosonic final states, respectively.
The reheating process can then be solved for each of those processes separately, but using the parametrization of Ref.~\cite{Garcia:2020wiy}, the inflaton decay width can be parametrized as
\begin{equation}
\label{eq:generic gammaphi}
    \Gamma_{\phi}=\gamma_{\phi}\left(\frac{\rho_{\phi}}{M_{P}^{4}}\right)^{l}\,.
\end{equation}
Now the details of the inflaton channel are contained in the parameters $\gamma_{\phi}$ and $l$. We specify the proper correspondence between the channels from Eq.~\eqref{eq:inflaton channels} and the decay width in App.~\ref{app:inflaton} along with a reminder of the inflaton dynamics during reheating. Using this width and considering a negligible radiation density at the end of inflation, the temperature will quickly peak at a maximum value $T_{max}$ before redshifting and achieving reheating once the condition $\rho_{R}(a_{RH})=\rho_{\phi}(a_{RH})$ is met\footnote{Note that this approach assumes instantaneous thermalization}. For our purpose of thermal dark matter production, we will consider the evolution starting from the radiation peak at a redshift denoted $a_{max}$ before a generic temperature scaling $T\sim a^{-\xi}$. In this framework, the solution of the system of Boltzmann equations using Eq.~\eqref{eq:generic gammaphi} gives the generic correspondence~\cite{Garcia:2020wiy}
\begin{align}
    \xi &\sim \frac{3k+6kl}{4k+8}\,,\\
    \frac{a_{max}}{a_{end}} &= \left(\frac{4k+8}{3k+6kl}\right)^{\frac{k+2}{k+8-6kl}}\,,
\end{align}
\noi
 valid as long as $8+k-6kl>0$. We will restrict ourselves to this case in the following, specializing to $k=2$ and $k=4$. In this case the temperature profile during reheating generally solves to \cite{Garcia:2020wiy}
\begin{equation}
\label{eq:reheating profile}
    \rho_{R} =\frac{2k}{k+8-6kl}\frac{\gamma_{\phi}}{H_{end}}\frac{\rho_{end}^{l+1}}{M_{P}^{4l}}\left(\frac{a_{end}}{a}\right)^{4}\left[\left(\frac{a}{a_{end}}\right)^{\frac{k+8-6kl}{k+2}}-1\right]
\end{equation}
\noi
 Note that throughout this work we will fix the reheating temperature and consider compatible values of $\xi$ corresponding to the introduced models while parametrizing everything from the reheating point. 

\section{Parameter space for relativistic freeze-out}
\label{sec:UV_IR_UFO}
\subsection{Generalities}

The main condition for UFO to be operative is for the dark matter to enter and exit thermal equilibrium using the criterion from Eq.~\eqref{eq:thermal eq}, while ensuring that the departure from equilibrium occurs while the dark matter is (ultra)relativistic. Assuming the dark matter to be relativistic, its interaction rate from Eq.~(\ref{eq:generic sigmav}) can be expressed
\begin{equation}
\label{eq:relativistic interaction rate}
    \Gamma = \frac{g_{\chi}\zeta(3)}{\pi^{2}}\frac{T^{n+3}}{\Lambda^{n+2}}\,.
\end{equation}
\noi
Here, $\chi$ denotes the dark matter candidate for which we will take $g_{\chi} = 1$ corresponding to a scalar singlet dark matter.\footnote{This assumption is not critical for our results, which can be readily generalized to fermionic dark matter.} Next, we need to compute the Hubble rate, which can be expressed in terms of the reheating temperature
\begin{equation}
\label{eq:generic hubble}
H(T)=\frac{\sqrt{\alpha}T_{RH}^{2}}{\sqrt{3}M_{P}}\left(\frac{T}{T_{RH}}\right)^{\frac{3k}{\xi(k+2)}}\,.
\end{equation}
\noi
With $\alpha = \pi^{2}g_{\rho}/30$ coming from the radiation energy density at reheating time ($\rho_{RH} = \alpha T_{RH}^{4}$). Requiring dark matter to enter and later on relativistically decouple from equilibrium then imposes a condition on the temperature scaling of both quantities. Indeed, decoupling can only happen if the interaction rate has a steeper temperature dependence than the Hubble rate, implying the generic condition
\begin{equation}
\label{eq:UFO condition}
    n>\frac{3k-3\xi(k+2)}{\xi(k+2)} =\frac{1-6l}{1+2l}\,.
\end{equation}
\noi
This matches the condition previously found in Ref.~\cite{Henrich:2025gsd} for $l = 1/2-1/k$,
namely $n>(3-k)/(k-1)$. If one evaluates Eq.~\eqref{eq:UFO condition} for the two other reheating channels, this condition reduces to $n>2k-3$ for $\phi\rightarrow bb$ and to $n>2k/3-3$ for $\phi\phi\rightarrow bb$. These results are relevant if one considers each reheating channel to be the only source of radiation. This is, however, not always the case, as one can imagine a reheating with 2 different inflaton couplings or a radiation source through graviton exchange~\cite{Clery:2021bwz} ($\xi = 1$). These modifications of the temperature evolution will be important as they may alter how the dark matter will exit thermal equilibrium. The UFO condition for $n$ can then be sumarized in Tab.~\ref{tab:critical_n}

\begin{table}[h]
\centering
\renewcommand{\arraystretch}{1.4}
\begin{tabular}{|c|c|c|c|}
\hline
 & $k=2$ & $k=4$ & $k=6$ \\
\hline
$\phi\rightarrow f\bar{f}$ & $1$ & $-\frac{1}{3}$ & $-\frac{3}{5}$ \\
\hline
$\phi\rightarrow bb$ & $1$ & $5$ & $9$ \\
\hline
$\phi\phi\rightarrow bb$ & $\times$ & $-\frac{1}{3}$ & $1$ \\
\hline
$\xi=1$ & $-\frac{3}{2}$ & $-1$ & $-\frac{3}{4}$\\
\hline
\end{tabular}
\caption{Minimal value of $n$ required for UFO to exit thermal equilibrium.}
\label{tab:critical_n}
\end{table}
\noi
The channel $\phi\phi\rightarrow bb$ is omitted for $k=2$ since it cannot induce proper 
reheating ($\xi=1$). In this case, the remaining channels yield identical results, as the 
decay width is constant regardless of the channel, recovering the standard scaling 
$T\propto a^{-3/8}$. A similar degeneracy occurs for $k=4$ between the fermionic decay and 
bosonic scattering channels, both of which produce the same temperature evolution 
$T\sim a^{-3/4}$. Among all channels, the fermionic decay proves to be the most efficient for 
UFO, while the bosonic decay is the most constrained, owing to its systematically shallower 
temperature scaling. As we will see, however, this more gradual scaling becomes advantageous 
when additional radiation production is taken into account. Finally, in the last line, we compute the bound for $\xi=1$ corresponding to the case where an abrupt initial radiation production would dominate before the reheating channel dominates. We note that heavy mediator interactions with $n=2$ (which are paradigmatic UFO interactions \cite{Henrich:2025pca,Henrich:2026tox}) are compatible with all reheating channels we consider, apart from the high $k$ $\phi\rightarrow bb$ channel.

\subsection{Limit between UV and IR dominated freeze-out}

Before turning to more general scenarios, we first establish the key condition 
distinguishing UV-like from IR-like production. A remarkable feature of UFO is its ability 
to induce a freeze-in like phase once the dark matter exits thermal equilibrium. 
This effect arises from the monomial temperature dependence of both the interaction rate 
and the Hubble rate, which causes decoupling to proceed more gradually than in the standard 
WIMP case, where the Boltzmann suppression drives a sharp departure from equilibrium. Relativistic decoupling during a period of entropy production (such as reheating or early matter domination) is also crucial for a post-UFO freeze-in phase. When UFO occurs during radiation domination, no additional freeze-in phase occurs since the comoving radiation number remains constant. To 
make this explicit, we work from the Boltzmann equation perspective and express it in terms of the yield 
$Y_{\chi} = n_{\chi} a^3$.
\begin{equation}
\label{eq:Boltzmann Yield}
    \frac{dY_{\chi}}{da} = \frac{\langle\sigma v\rangle a^{2}}{H}(n_{eq}^{2}-n_{\chi}^{2})\,.
\end{equation}

\begin{figure}
    \centering
    \includegraphics[width=\linewidth]{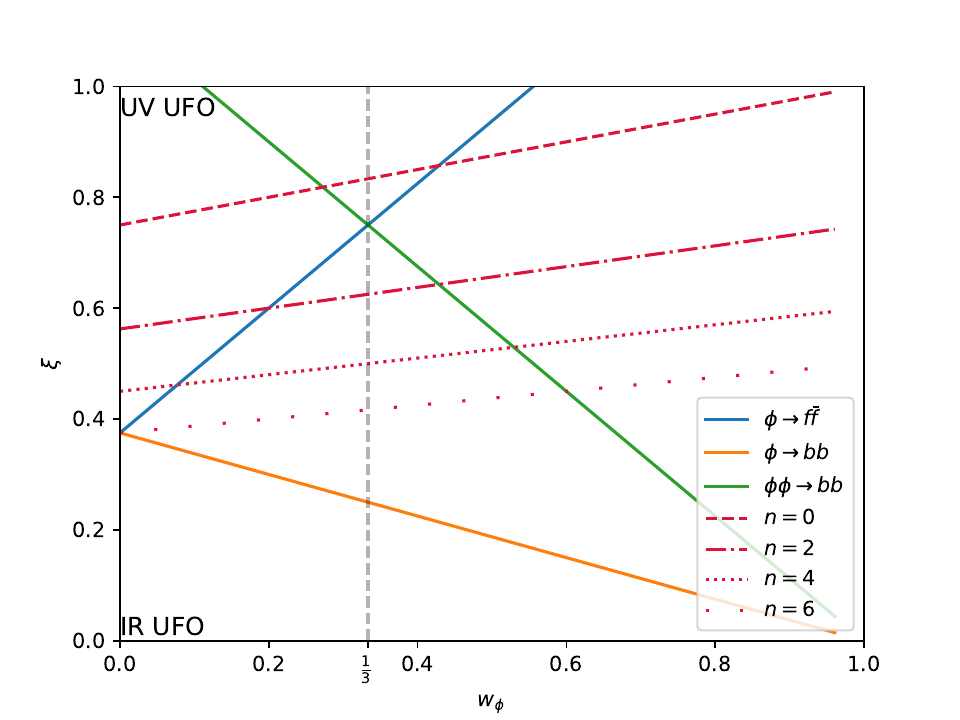}
    \caption{\justifying Parameter space between UV and IR UFO in the $w_\phi$, $\xi$ plane for different temperature scaling of the thermally averaged cross section. For each $n$ value the part above the crimson line correspond to UV UFO while the part below correspond to IR UFO. We show the various reheating channels with the colored full lines. The cases $k=2,4$ on which we will focus respectively correspond to $w_{\phi} = 0,\frac{1}{3}$.}
    \label{fig:UV_IR}
\end{figure}

\noi
Once relativistic decoupling happens, we keep only the equilibrium term on the right hand side of Eq.~\eqref{eq:Boltzmann Yield} as the entropy injection will slow the decrease of the equilibrium number density compared to the relic implying $n_{eq}\gg n_{\chi}$\footnote{Note that this should not be done in the case $\xi=1$ as the Yield and the equilibrium Yield will evolve in the same way canceling the source term.}. Note that this is the opposite of what typically happens for WIMP-like freeze-out, where the annihilation term proportional to $n_\chi^2$ drives the dynamics after freeze-out. Using the parametrization for the thermally averaged cross section, the relativistic number density, and the temperature profile during reheating, the source term scaling then goes like
\begin{equation}
    \frac{dY_{\chi}}{da}\propto a^{\frac{5k+4-\xi(n+6)(k+2)}{k+2}}\,.
\end{equation}
\noi
This polynomial scaling will then exhibit the two different standard behaviors depending on if the source term decreases faster or slower compared to $\sim a^{-1}$. If the source term decrease faster than $\sim a^{-1}$, UFO will be dominated by the production at early time near the decoupling (small $a$)  which we will refer to as UV UFO while in the other case the production will be dominated by the late time (big $a$) which we will referred to as IR UFO \footnote{For UV UFO, the characteristic UV scale which will dictate the abundance is $T_{\rm FO}$, while the characteristic scale for IR UFO is typically max($T_{\rm RH},m_\chi$).}. The limiting case between these two regimes then sets a critical value of $n$ given by\footnote{This result reproduces the one from \cite{Deka:2026psw} as visible in Fig.~\ref{fig:UV_IR}.}
\begin{equation}
    n_{c} = 6\frac{k+1-\xi(k+2)}{\xi(k+2)} = \frac{2k+8-12kl}{k+2kl}\,.
\end{equation}

\noi
Recasting this constraint as a link between $k$ and $\xi$ then allows one to delimit the parameter space between UV and IR UFO depending on the inflaton equation of state and the temperature scaling $\xi$. We display this limit in Fig.~\ref{fig:UV_IR}. The latter clearly shows that for any reheating channel UFO will be IR like for a quadratic potential and UV like for quartic potentials, except for the case $\phi\rightarrow bb$ if $n\geq 2$. Note that we did not include here the limits related to consistent reheating as our purpose is to study how a non trivial evolution of the temperature, i.e. having different eras of temperature evolution and different values of $\xi$, can modify the UFO parameter space. More specifically we will specialize in the minimal case where gravity will source radiation. In this case if the reheating temperature is sufficiently low, UFO is not guaranteed to happen during the standard $T\sim a^{-\xi}$ era and the gravitationally generated bath might become relevant as it can change the UFO nature from IR like to UV like as we shall now demonstrate.

\section{Gravitationally produced radiation and UFO}
\label{sec:GUFO}
In recent years it was noticed that high scale inflation should be accompanied by a minimal radiation production through gravitational interactions \cite{Clery:2021bwz}. This production stems from taking Einstein's General Relativity as an effective theory below the Planck scale. At tree level, it then predicts an energy transfer from the inflaton to radiation, implying a minimal temperature for the radiation bath around $\sim 10^{12}\,\mathrm{GeV}$ soon after the end of inflation. A remarkable feature of this production is its independance on the reheating channel; the exact value is fixed solely by the inflaton potential.
This minimal production is then directly relevant for Freeze-in dark matter as the graviton mediated interaction can modify the production from the thermal bath\footnote{Note that this change requires a sufficiently low reheating temperature.} and add a contribution $\phi\phi\rightarrow \chi\chi$  which can be efficient when the inflaton starts to oscillate. That is because this channel is sourced by the inflaton energy density at the end of inflation, $\rho_{\phi}(a_{\rm end})$, which is quite high ($\approx 10^{62}$ GeV$^4$) for high-scale inflation~\cite{Clery:2021bwz,Clery:2022wib,Henrich:2024rux}. In the WIMP case, however, this effect is less dramatic than for freeze-in. Thermal equilibrium erases the dependence on the initial conditions, and this effect could at most modify the dark matter dilution if freeze-out happens sufficiently early \cite{Bernal:2022wck}.
The case of UFO, though, is singular as the temperature evolution is important both to reach freeze-out and determine the UV/IR behavior of UFO as discussed in the previous section. From now on, we focus on a scenario in which reheating occurs through one of the channels presented in Eq.~\eqref{eq:inflaton channels}, and we include the unavoidable, gravitationally produced radiation. In this scenario, the temperature profile will now receive an additional contribution noted $\rho_{R}^{h}$ which solves \cite{Clery:2021bwz}
\begin{equation}
\label{eq:gravitational rad}
    \frac{d\rho_{R}^{h}}{dt}+4H\rho_{R}^{h} = N\frac{\rho_{\phi}^{2}\omega}{16\pi^{4}M_{P}^{4}}\Sigma_{k}^{h}\,,
\end{equation}
with $\Sigma_{k}^{h}$ a sum over the inflaton modes which evaluates according to the following table~\cite{Clery:2021bwz}.
\begin{table}[h]
\centering
\begin{tabular}{|c|c|c|c|}
\hline
 & $k=2$ & $k=4$ & $k=6$ \\
\hline
$\Sigma_k^{h}$  & $\frac{1}{8}$ & $0.126$ & $0.124$ \\
\hline
\end{tabular}
\end{table}

\noi
Note that we will take $N=4$ being the number of real scalar degrees of  freedom in the standard model\footnote{ The contribution from the graviton portal needs to be evaluated coherently with the considered model and the potential non minimal couplings to gravity~\cite{Clery:2022wib} which we do not consider here. For example in supersymetric extensions $N$ will be bigger corresponding to the number of scalar degrees of freedom. }. $\omega$ is the inflaton oscillation frequency. Details regarding the inflaton oscillations are given in appendix~\ref{app:inflaton}. The extra radiation contribution  can then be integrated to get \cite{Clery:2021bwz} 
\begin{eqnarray}
\label{eq:grav rad profile}
    \rho_{R}^{h} = N\frac{\sqrt{3}M_{P}^{4}\gamma_{k}\Sigma_{k}^{h}}{16\pi}\left(\frac{\rho_{end}}{M_{P}^{4}}\right)^{\frac{2k-1}{k}}\frac{k+2}{8k-14}\\ \times\left[\left(\frac{a_{end}}{a}\right)^{4}-\left(\frac{a_{end}}{a}\right)^{\frac{12k-6}{k+2}}\right]\nonumber\,,
\end{eqnarray}
where the factor $\gamma_{k}$ can be expressed in terms of the potential coupling and power as
\begin{equation}
    \gamma_{k} = \sqrt{\frac{\pi}{2}}k\frac{\Gamma(\frac{1}{2}+\frac{1}{k})}{\Gamma(\frac{1}{k})}\lambda^{\frac{1}{k}}\,.
\end{equation}
\noi
In the end, this contribution peaks soon after the end of inflation, like the standard reheating production, giving the following maximum radiation density \cite{Clery:2021bwz}
\begin{equation}
    \rho_{R,max}^{h} = N\frac{\sqrt{3}M_{P}^{4}\gamma_{k}\Sigma_{k}^{h}}{16\pi}\left(\frac{\rho_{end}}{M_{P}^{4}}\right)^{\frac{2k-1}{k}}\frac{k+2}{12k-6}\left(\frac{2k+4}{6k-3}\right)^{\frac{2k+4}{4k-7}}\,,
\end{equation}

\noi
at a maximum redshift
\begin{equation}
    \frac{a_{end}}{a_{max,h}}=\left(\frac{2k+4}{6k-3}\right) ^{\frac{k+2}{8k-14}}\,.
\end{equation}
\noi
 As a consequence, this production is a minimal unavoidable contribution, suggesting that, for a sufficiently low reheating temperature, it should dominate at least around $a_{max}$. The existence of the gravitationally generated radiation bath then imposes an early temperature scaling $T\sim a^{-1}$ prior to the period with scaling $T\sim a^{-\xi}$, where the latter leads to proper reheating. Requiring this initial period of domination by the gravitationally generated bath, therefore, places an upper bound on the allowed reheating temperature
\begin{equation}
\label{eq:hdomination}
    T_{RH}<\frac{(\rho_{R,max}^{h})^{\frac{3k}{12k-8\xi(k+2)}}}{\alpha^{\frac{1}{4}}\rho_{end}^{\frac{\xi(k+2)}{6k-4\xi(k+2)}}}\left(\frac{a_{max,h}}{a_{end}}\right)^{\frac{3\xi k}{3k-2\xi(k+2)}}\,.
\end{equation}

As long as the reheating temperature satisfies this bound, the temperature profile will first evolve with $T\sim a^{-1}$ until the standard reheating contribution starts to dominate at a scale factor $a_{\times}$ corresponding to a temperature $T_{\times}$. To compute it we use Eq.~\eqref{eq:reheating profile} and Eq.~\eqref{eq:grav rad profile} in the limit $a\gg a_{end}$ noting that $\rho_{R}^{h}\sim\rho_{R,e}^{h}(a_{end}/a)^{4}$ which gives for the crossing redshift $a_{\times}$
\begin{align}
    \frac{a_{\times}}{a_{end}} &= \left(\frac{\rho_{R,e}^{h}}{\rho_{RH}}\right)^{\frac{1}{4(1-\xi)}}\left(\frac{\rho_{RH}}{\rho_{end}}\right)^{\frac{\xi(k+2)}{6k(1-\xi)}}\,, \label{aCross}\\
    \frac{a_{\times}}{a_{RH}} &= \left(\frac{\rho_{R,e}^{h}}{\rho_{RH}}\right)^{\frac{1}{4(1-\xi)}}\left(\frac{\rho_{RH}}{\rho_{end}}\right)^{\frac{(k+2)}{6k(1-\xi)}}\,,
\end{align}
\noi
implying, hence, for the ratio between $T_{\times}$ and $T_{RH}$ that
\begin{equation}
    \frac{T_{\times}}{T_{RH}} =\left(\frac{a_{RH}}{a_{\times}}\right)^{\xi} = \frac{\alpha^{\frac{\xi(k-4)}{12k(1-\xi)}}\rho_{end}^{\frac{\xi(k+2)}{6k(1-\xi)}}}{(\rho_{R,e}^{h})^{\frac{\xi}{4(1-\xi)}}}T_{RH}^{\frac{\xi(k-4)}{3k(1-\xi)}} := \mathcal{T}T_{RH}^{\frac{\xi(k-4)}{3k(1-\xi)}}\,.\label{Tcross}
\end{equation}
\noi
As a specific example, for the familiar case where $k=2$ and $T\propto a^{-3/8}$, Eqs.~(\ref{aCross}) and (\ref{Tcross}) give
\begin{equation}
\frac{a_{\times}}{a_{end}}\propto \left(\frac{T_{max,h}^{8}}{T_{RH}^{4} \rho_{end}}\right)^{1/5}, \hspace{3mm} \frac{T_{\times}}{T_{\rm RH}}\propto \left(\frac{\rho_{end}}{T_{max,h}^{3}T_{RH}}\right)^{1/5}\,.
\end{equation}
\noi
 Expressing the crossing temperature in a more direct form gives
\begin{equation}
\label{eq:Tcross}
    T_{\times} = \mathcal{T}T_{RH}^{\frac{\xi(k-4)+3k(1-\xi)}{3k(1-\xi)}}\,\,,
\end{equation}

\begin{figure}
    \centering
    \includegraphics[width=\linewidth]{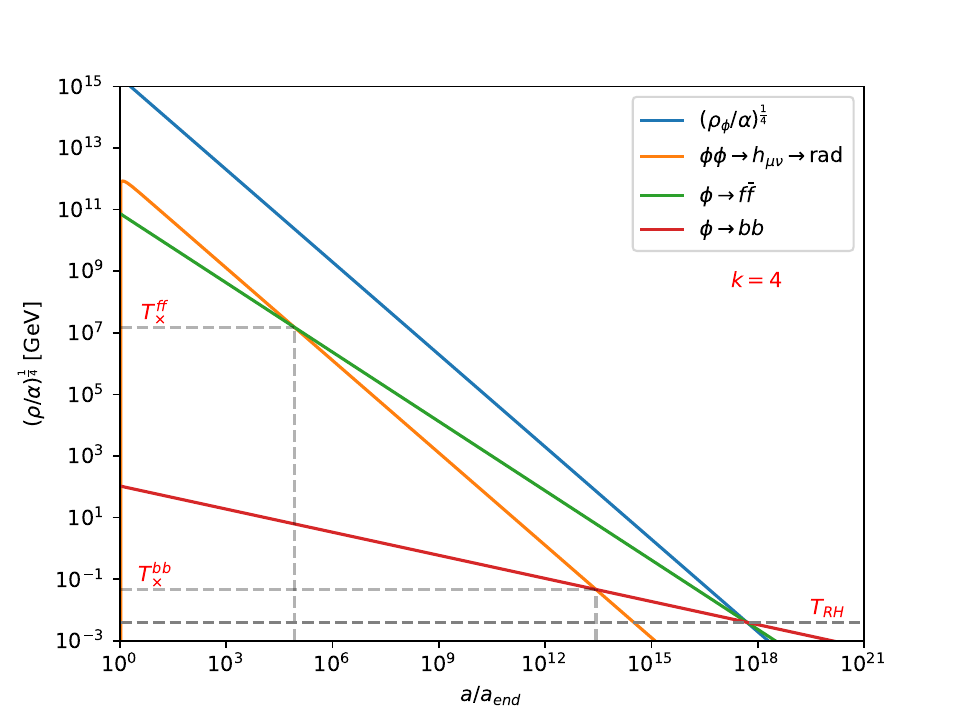}
    \caption{\justifying Temperature profile during reheating considering the reheating channel $\phi\rightarrow bb$ (red) and $\phi\rightarrow f\bar{f}$ (green) and the graviton mediated production (orange) for $k=4$. We highlight explicitly the crossing times and temperatures in each case with dashed lines using Eq.~\eqref{aCross} and Eq.~\eqref{eq:Tcross}. We fixed $T_{RH} = 4\,\mathrm{MeV}$ for readability.}
    \label{fig:temperature profile}
\end{figure}

\noindent where we have introduced above the quantity $\mathcal{T}$ for simplicity\footnote{Note that $\mathcal{T}^{\frac{1-\xi}{\xi}}$ does not depend on the reheating channel.}. 
This constant of dimension $\frac{\xi}{1-\xi}\frac{4-k}{3k}$ is fixed once one chooses a specific model, and it allows one to conveniently express the Hubble scale for $a<a_{\times}$
\begin{equation}
\label{eq:Hgrav}
    H(T) = \frac{\sqrt{\alpha}}{\sqrt{3}M_{P}}\mathcal{T}^{\frac{3k(1-\xi)}{\xi(k+2)}}T^{\frac{3k}{k+2}}\,.
\end{equation}
\noi
Using Eq.~\eqref{eq:thermal eq}, Eq.~\eqref{eq:generic hubble}, and Eq.~\eqref{eq:Hgrav}, we can derive the generic ultrarelativistic freeze-out temperatures during each radiation phase
\begin{align}
\label{eq:genericTforeh}
    T_{FO,reh}=T_{RH}\left(\frac{\sqrt{3\alpha}\pi^{2}(1+w_\phi)}{2g_{\chi}\zeta(3)}\frac{\Lambda^{n+2}}{M_{P}T_{RH}^{n+1}}\right)^{\frac{\xi(k+2)}{\xi(k+2)(n+3)-3k}}\,,\\
\label{eq:genericTfoh}
    T_{FO,h} = \left(\frac{\sqrt{3\alpha}\pi^{2}(1+w_\phi)}{2g_{\chi}\zeta(3)}\frac{\mathcal{T}^{\frac{3k(1-\xi)}{\xi(k+2)}}\Lambda^{n+2}}{M_{P}}\right)^{\frac{k+2}{(k+2)(n+3)-3k}}\,,
\end{align}
which correspond, respectively, to the standard UFO and to UFO from the gravitational bath, which we will refer to as GUFO from now on.
 The key feature here is the behavior with respect to the reheating temperature. For freeze-out from the reheating bath (which is sourced by the rate given in Eq.~(\ref{eq:generic gammaphi})), there is a direct dependence on the reheating temperature, leading to a minimum value of $T_{RH}$ for freeze-out to occur.  This feature also constrains the parameter space to enter thermal equilibrium, which is a key limitation of standard UFO \cite{Henrich:2025gsd}. On the other case, if freeze-out happens from the gravitationally produced bath, the freeze-out temperature is independent on $T_{RH}$. This feature leads to a maximum reheating temperature necessary to satisfy $T_{FO,h}>T_{\times}$\footnote{Note that requiring $T_{FO,h}>T_{\times}$ and $T_{FO,reh}<T_{\times}$ gives equivalent constraints for $\Lambda$.} as the latter grows with respect to $T_{RH}$. This is one of our main results, since if one considers this extra radiation bath in the context of a small enough reheating temperature, there will be an unavoidable gravitational UFO for sufficiently high BSM scale $\Lambda$. Standard UFO will remain possible, but as outlined in Tab.~\ref{tab:critical_n}, it is well captured by the analysis done in Ref.~\cite{Henrich:2025gsd} as long as $T_{FO,reh}<T_{\times}$. We should stress that this last constraint is important, as GUFO may not have the same phenomenology as UFO for all cases.

 For the remainder of this work, we focus on GUFO and require the dark matter decoupling temperature $T_{FO,h}$ to be above the crossing temperature $T_{\times}$ while being sure not to enter thermal equilibrium again when the temperature profile changes from $T\sim a^{-1}$ to $T\sim a^{-\xi}$. This last requirement is guaranteed in the case where $n$ satisfies the condition from Eq.~\eqref{eq:UFO condition} with the appropriate $\xi$ as it is a stronger condition than the one in the gravitationally produced radiation (i.e. the UFO condition is always more stringent than the GUFO one). The picture is slightly more complex in the other case, where only GUFO is possible. In this case, as the particle would normally not be able to freeze-out ultrarelativistically, it might enter back in thermal equilibrium if reheating takes too long after the crossing \footnote{ We note that there is not a problem \textit{a priori} with re-entering equilibrium after undergoing UFO or GUFO, as the dark matter may freeze-out non-relativistically after re-entering equilibrium, and in principle could still achieve the correct abundance. However, since we focus on UFO and GUFO rather than WIMP-like freeze-out here, we omit these cases from our analysis.}. This can be translated into a limit for the duration of this phase arising from requiring no thermal equilibrium at reheating time, which in turn gives a limit on the GUFO freeze-out temperature 
\begin{equation}
\label{eq:rethermal condition}
    T_{FO,h}>\underbrace{T_{\times}\left(\frac{T_{RH}}{T_{\times}}\right)^{\frac{\xi(k+2)(n+3)-3k}{\xi(k+2)(n+3)-3k\xi}}}_{T_{th}}\,,
\end{equation}
where here the previous intuition that GUFO can enter back in thermal equilibrium is manifest in the exponent. The denominator is always positive as $n>-\frac{6}{k+2}$ for GUFO to happen, while the numerator is positive only if Eq.~\eqref{eq:UFO condition} is satisfied, i.e. if standard UFO is accessible. Then, since we should always satisfy $T_{FO,h}>T_{\times}$ and $T_{\times}>T_{RH}$ the condition in Eq.~\eqref{eq:rethermal condition} is relevant only when standard UFO is not accessible. The proper condition for GUFO to be accessible can then be written
\begin{equation}
\label{eq:GUFO condition}
    T_{FO,h} >  \mathrm{max}(T_{\times},T_{th})\,,
\end{equation}
which we evaluate numerically as a constraint on the BSM scale in the following form

\begin{figure*}[t]
\centering

\includegraphics[width=0.5\textwidth]{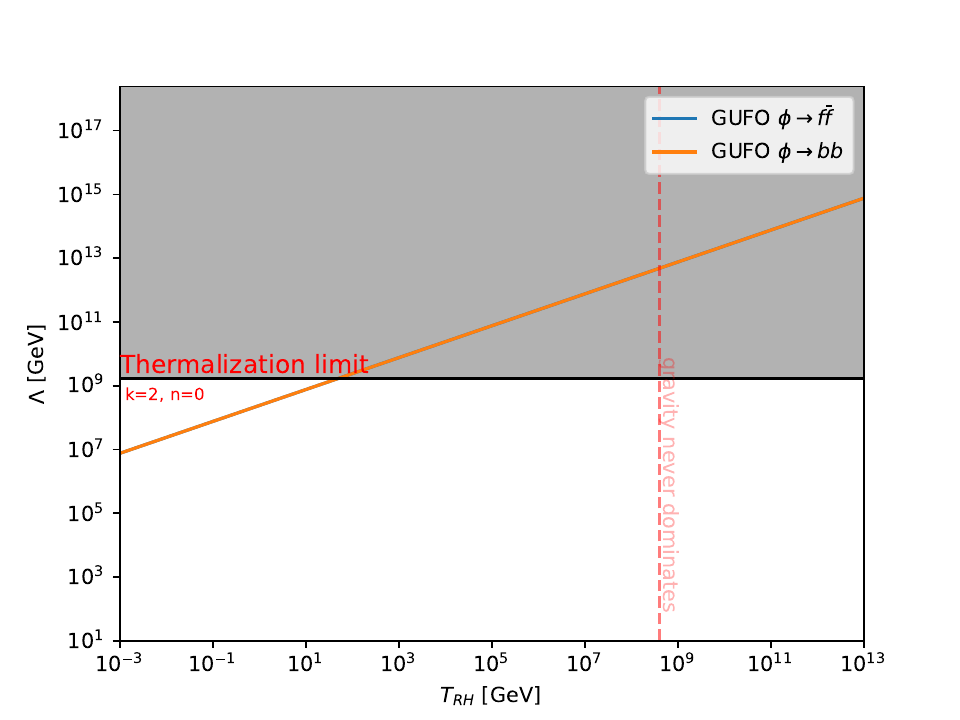}%
\includegraphics[width=0.5\textwidth]{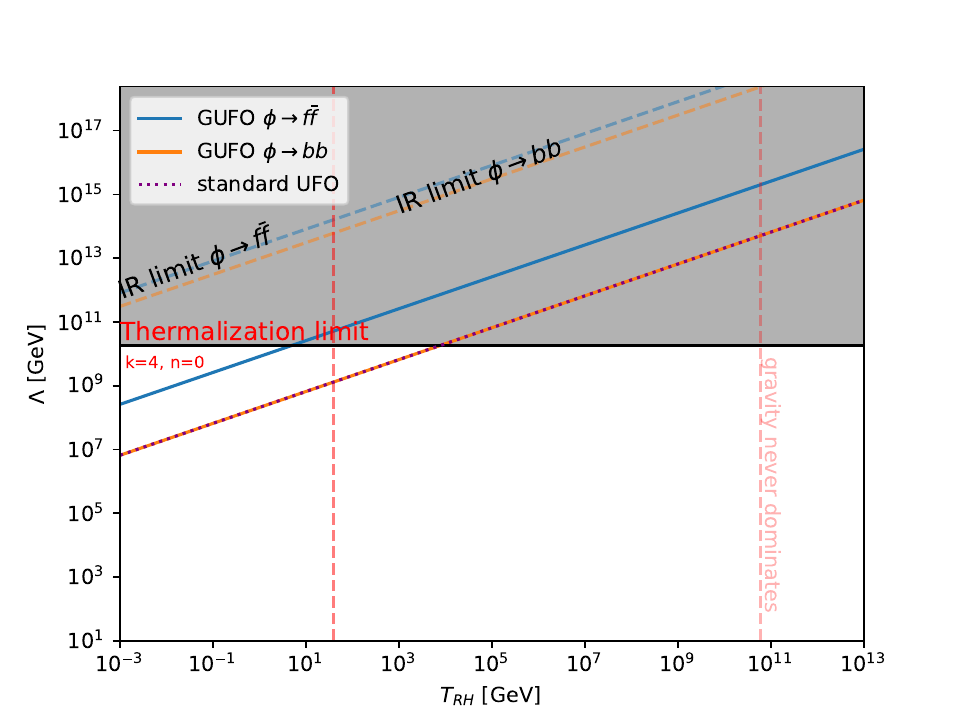}

\includegraphics[width=0.5\textwidth]{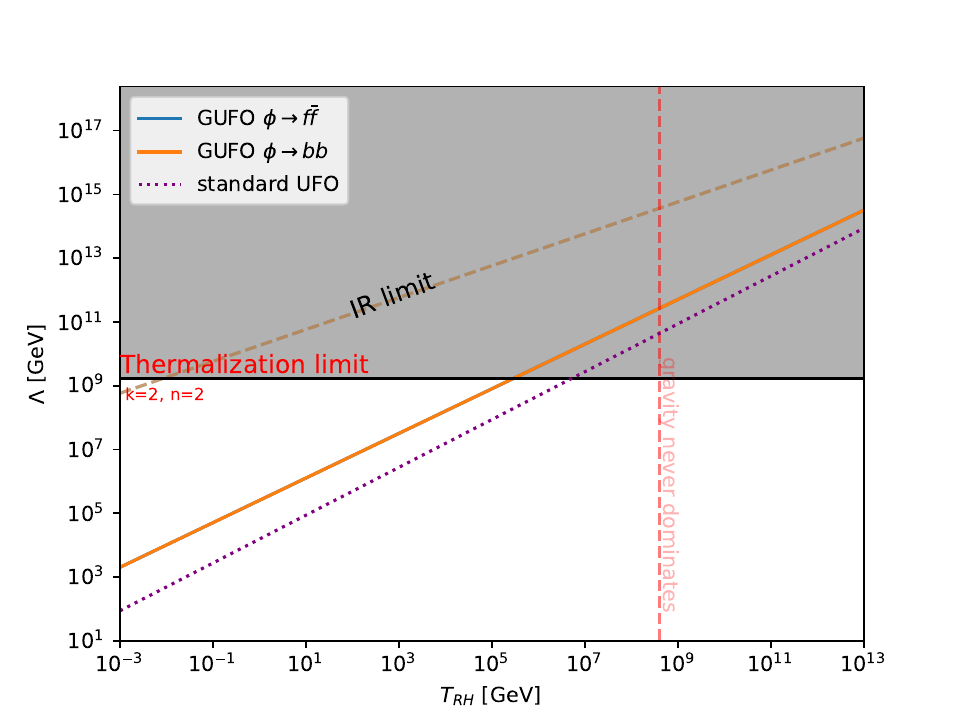}%
\includegraphics[width=0.5\textwidth]{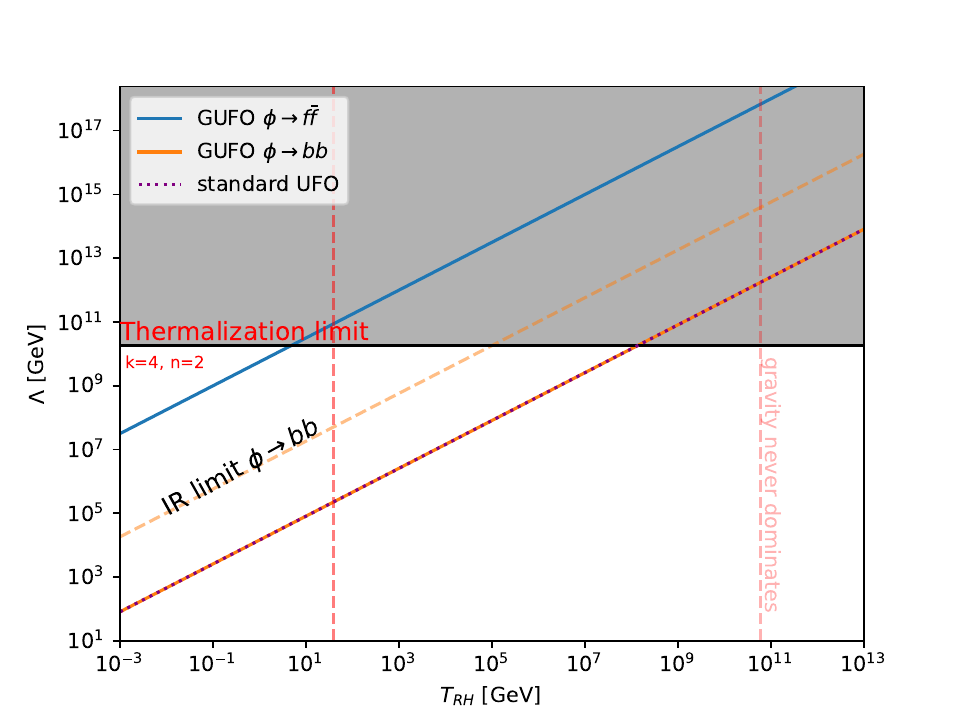}

\caption{\justifying Parameter space for GUFO to be accessible for the 4 possibilities $n=0,2$ and $k=2,4$. The left (right) column corresponds to $k=2 (4)$, while the upper (lower) line depicts the case for $n=0  (2)$. On top of the GUFO limits coming from Eq.~\eqref{eq:lambda constraint GUFO} and given in full lines, the standard UFO condition is displayed in dotted line when relevant. In addition, we include the limit for the dark matter to thermalize with the gravitational bath by imposing $T_{FO,h}<\Lambda$ (no thermalization with the gravitational bath is possible in the gray region) and the limit on $T_{RH}$ so the gravitationally produced radiation will dominate the radiation energy density from Eq.~\eqref{eq:hdomination} (for $T_{RH}$ values greater than those indicated by the vertical red dashed lines, the gravitational bath will never be relevant). We only display the reheating channels $\phi\rightarrow f\bar{f}$ and $\phi\rightarrow bb$ as the scattering channel is either irrelevant or equivalent to the fermionic case for $k=4$. For the case $k=2\,,n=0$ the IR limit lies above the range of the plot.}
\label{fig:parameter_space}
\end{figure*}

\be
\label{eq:lambda constraint GUFO}
 \Lambda >\left(\frac{2 g_{\chi}\zeta(3)M_{P}}{\sqrt{3\alpha}\pi^{2}(1+w_{\phi})\mathcal{T}^{\frac{3k(1-\xi)}{\xi(k+2)}}}\right)^{\frac{1}{n+2}}\mathrm{max}(T_{\times},T_{th})^{m} \nn
\ee
with $m = \frac{(k+2)(n+3)-3k}{(k+2)(n+2)}$.
We display the various cases in Fig.~\ref{fig:parameter_space} for the 4 combinations $k=2,4$ and $n=0,2$\footnote{Note that we do not display the case $\phi\phi\rightarrow bb$ as it is irrelevant for $k=2$ and equivalent to $\phi\rightarrow f\bar{f}$ for $k=4$.}. As one can see from the figure, depending on the reheating scenario, GUFO can cover a large fraction of the parameter space compared to standard UFO especially for $n=2$, $k=2$. Moreover, its main impact is to open up otherwise forbidden scenarios, such as the bosonic decay channel $\phi \rightarrow bb$ for $k = 4$, as well as the case $k = 2$, $n = 0$.
We display in dashed lines the limit between the UV and IR dominated GUFO when relevant. This limit is computed in the next section where we will comment on the various cases. For context we display the condition for standard UFO when relevant in dotted violet line and the thermalization limit for GUFO in black requiring  $T_{FO,h}<\Lambda$. 

Finally we should comment on the coincidence of the standard UFO condition and the GUFO condition from Eq.~\eqref{eq:lambda constraint GUFO}. This may seems strange at first sight but formally one can inject the set of parameters $k=4,\,\xi= 1/4$ in Eq.~\eqref{eq:genericTfoh} and Eq.~\eqref{eq:rethermal condition} and the GUFO condition presented in Eq.~\eqref{eq:GUFO condition} then boils down to the standard UFO condition $T_{FO,reh}>T_{RH}$ meaning that for $k=4$ with the bosonic decay channel, only GUFO is possible and is even more accessible than the same scenario with the two other reheating channels. This can be understood quite easily. The main limitation on GUFO to be compatible with a new interaction type is given by Eq.~\eqref{eq:rethermal condition}, which ensures that the dark matter will refrain from re-entering thermal equilibrium when the temperature profile goes back to $T\sim a^{-\xi}$. This is fundamentally a limit on the duration of this period.  The channel $\phi\rightarrow bb$ has the slowest decrease of the temperature therefore limiting the possibility for dark matter to rereach thermal equilibrium implying a bigger parameter space for this case.
Note that here we did not discuss the important condition that the freeze-out temperature should be above the particle mass to limit the number of parameters in  this section. However, this extra condition should be applied in order for the analysis to remain consistent and we will take it into account later on for the correct dark matter relic abundance production as the mass will inevitably enter the relations. Now that we understand better the parameter space and the relevance of GUFO to describe relativistic decoupling during reheating we will move on to producing the correct relic abundance. 

\section{Dark matter abundance}
\label{sec:relic abundance}
\subsection{Computing the Yield}

As we just discussed, GUFO is an unavoidable process that will modify the standard UFO in a wide portion of the parameter space. We now need to link the BSM scale, the mass and the reheating temperature to the relic abundance. Starting again from the Boltzmann equation presented in Eq.~\eqref{eq:Boltzmann Yield}, the total Yield will be the sum of two terms
\begin{align}
\label{eq:dark matter production}
    Y_{\chi}(a_{RH}) &= Y_{FO} + \int_{a_{\times}}^{a_{RH}}\frac{n_{eq}^{2}\langle\sigma v\rangle a^{2}}{H}da\,,\nonumber \\
    &=Y_{FO}+Y_{reh}\,.
\end{align}
The first term $Y_{FO}$ corresponds to the equilibrium yield at the time of freeze-out, and the second $Y_{reh}$ is the production happening during the later stage of reheating, where $T\sim a^{-\xi}$. At this stage, one might wonder why we do not consider a potential production between $a_{FO,h}$ and $a_{\times}$. The answer follows from the temperature scaling $T\sim a^{-1}$. Directly after freeze-out, the number density is equal to the equilibrium number density, so the source term will be null, implying the standard scaling for the number density, $n\propto a^{-3}$. This is the same as the equilibrium number density since $n_{eq}\sim T^{3}\sim a^{-3}$. Throughout this period, the Boltzmann equation is blocked, and no particles can be efficiently produced.   Once the temperature profile goes to $T\sim a^{-\xi}$, the equilibrium number density will redshift more slowly than the relic, allowing particle production again if the source term becomes big enough, like in standard UFO. In the end, the computation of the final abundance will in some sense be reduced to the standard UFO result with the subtlety that decoupling may occur significantly before the temperature scaling changes. Before moving to the potential IR production, let us compute the freeze-out Yield 
\begin{align}
    Y_{FO,h} &= n_{\chi}a_{FO,h}^{3}=\frac{g_{\chi}\zeta(3)}{\pi^{2}}T_{\times}^{3}\left(\frac{T_{RH}}{T_{\times}}\right)^{\frac{3}{\xi}}a_{RH}^{3},\nonumber\\
    \label{eq:YGUFO}
    &= \frac{g_{\chi}\zeta(3)}{\pi^{2}\mathcal{T}^{\frac{3(1-\xi)}{\xi}}}T_{RH}^{\frac{2k+4}{k}}a_{RH}^{3}\,,
\end{align}
 where the main feature is the independence with respect to $\Lambda$  because the intial phase, $T\sim a^{-1}$, cancels the dependence in $T_{FO,h}$. We should not be surprised by this feature as it is already the case for a standard relativistic freeze-out during radiation domination~\cite{Gershtein:1966gg,Cowsik:1972gh,Szalay:1974jta}. Fundamentally, this emerges since the Yield is frozen during this period, similarly to radiation domination due to the temperature profile. In that case, decoupling at two different moments will have no impact on the relic abundance, implying a dependence only on the reheating temperature. The latter corresponds to entropy injection during reheating. Importantly, the expression for $Y_{FO,h}$ in Eq.~(\ref{eq:YGUFO}) applies to GUFO, but not to standard UFO. For completeness, we also provide the following result for standard UFO, which applies when $T_{\times}>T_{FO}>T_{RH}$:

\begin{align}
    Y_{FO,reh}&= n_{\chi}(a_{FO})a_{FO,rh}^{3}=\frac{g_{\chi}\zeta(3)}{\pi^{2}}T_{FO,reh}^{3}a_{FO}^3 \nonumber \\
    \label{eq:YUFO}
    &= \frac{g_\chi \zeta(3)}{\pi^2} T_{RH}^3 a_{FO}^3 \times \nonumber \\
    &\!\left(\frac{\sqrt{3\alpha} \pi^2(1+w_\phi)}{2g_\chi\zeta(3)}\frac{\Lambda^{n+2}}{M_p T_{RH}^{n+1}}\right)^{\frac{3\xi(k+2)}{\xi(k+2)(n+3)-3k}} \,.
\end{align}

One important distinction between Eq.~(\ref{eq:YUFO}) and Eq.~(\ref{eq:YGUFO}) is the sensitivity to $\Lambda$. In standard UFO, the abundance at freeze-out is sensitive to both $\Lambda$ and $n$, but this is not the case for GUFO. Note that the subscripts $\{h, reh\}$ indicate the phase at which the dark matter freezes-out. However, since GUFO is the main new allowed feature according to Tab.~\ref{tab:critical_n}, from now on we will drop $h$ for clarity. Instead, we will express everything in terms of $Y_{FO,reh}$ when dealing with standard UFO. For the remainder of this section, we will only consider the case where the dark matter is never in equilibrium when $T<T_\times$, therefore specializing to the case of GUFO rather than standard UFO. Moving now to the second term of Eq.~\eqref{eq:dark matter production} , the source term can be formally integrated once the temperature profile changes

\begin{widetext}
    \begin{align}
        \label{eq:genericYreh}
    Y_{reh} &=\left(\frac{g_{\chi}\zeta(3)}{\pi^{2}}\right)^{2} \frac{T_{\times}^{n+6}a_{\times}^{\xi(n+6)}}{\Lambda^{n+2}H_{\times}a_{\times}^{\frac{3k}{k+2}}}\frac{k+2}{6(k+1)-\xi(n+6)(k+2)}\left[a_{RH}^{\frac{6(k+1)-\xi(n+6)(k+2)}{k+2}}-a_{\times}^{\frac{6(k+1)-\xi(n+6)(k+2)}{k+2}}\right]\,,
    \end{align}
\end{widetext}
with the two terms corresponding to the post-freeze-out contributions to the abundance for the UV or IR case of standard UFO. Contrary to the standard UFO case, however, this second phase starts only at $T_{\times}$ and not directly at $T_{FO,h}$, meaning that the source term will redishift between decoupling and crossing, hence reducing the production. Note that Eq.~(\ref{eq:genericYreh}) used a relativistic expression for the production rate $R_\chi\simeq\langle \sigma v\rangle n_{eq}^2$. As a result, it is only the correct formal result when $m_\chi \ll T_{RH}$. When $m_\chi \gtrsim T_{RH}$, one must instead use an upper limit of integration of $a=a_m$ (the scale factor at which $T=m_\chi$) for the relativistic production, which can then be used as an initial condition for non-relativistic production. In the UV case, one then expects the contribution from $Y_{reh}$ to be negligible relative to $Y_{FO}$ due to this extra redshift if $T_{FO,h}\gg T_{\times}$. Then, in this UV scenario one can write
\begin{align}
    Y_{reh,UV} &=\left(\frac{g_{\chi}\zeta(3)}{\pi^{2}}\right)^{2}\frac{k+2}{\xi(n+6)(k+2)-6(k+1)}\frac{T_{\times}^{n+6}a_{\times}^{3}}{\Lambda^{n+2}H_{\times}}\,,\nonumber\\
    &=\frac{k+2}{\xi(n+6)(k+2)-6(k+1)}\left.\frac{n_{eq}^{2}\langle\sigma v\rangle a^{3}}{H}\right|_{\times}\,,
\end{align}
neglecting the contribution from $a_{RH}$. To make the redshifting effect explicit in the above equation, we now rewrite this production in terms of $Y_{FO}$. It is straightforward to get
\begin{align}
\label{eq:source term redshift}
    \left.\frac{n_{eq}^{2}\langle\sigma v\rangle a^{3}}{H}\right|_{\times} &= \left.\frac{n_{eq}^{2}\langle\sigma v\rangle a^{3}}{H}\right|_{FO,h}\left(\frac{T_{\times}}{T_{FO,h}}\right)^{\frac{(n+6)(k+2)-6(k+1)}{k+2}}\,,\nonumber\\
    &=\frac{3(1+w_{\phi})}{2}Y_{FO}\left(\frac{T_{\times}}{T_{FO,h}}\right)^{\frac{(n+6)(k+2)-6(k+1)}{k+2}}\,,
\end{align}
from which we directly see that the $Y_{reh,UV}\ll Y_{FO}$ if $a_{\times}\gg a_{FO,h}$, i.e. if freeze-out happens deep in the gravitationally produced radiation domination. This is expected as a UV process is dominated by the early times since the source term is fastly decaying. In our case, the source term gets an additional damping from the redshift between $a_{FO,h}$ and $a_{\times}$.
Moving to the IR case, from Eq.~\eqref{eq:genericYreh}, the production will be dominated by the late time contribution, giving
\begin{equation}
    Y_{reh,IR} \simeq\left(\frac{g_{\chi}\zeta(3)}{\pi^{2}}\right)^{2}\frac{k+2}{6(k+1)-\xi(n+6)(k+2)}\frac{T_{RH}^{n+6}a_{RH}^{3}}{\Lambda^{n+2}H_{RH}}\,,
\end{equation}
reproducing exactly the standard IR UFO production. This is expected as IR production occurs in the later stages of reheating, where the standard model radiation bath temperature scales as $T\sim a^{-\xi}$  regardless of its previous behavior. 
We need now to evaluate this production compared to the initial freeze-out abundance as for the UV case. Indeed, after GUFO, the source term redshifts until $a=a_{\times}$, after which it begins to grow again. In a similar way to what we discussed for dark matter to stay outside of equilibrium during the $T\sim a^{-\xi}$ era, there should be a limit on the duration of this phase for this IR contribution to be relevant compared with the freeze-out one. To manifest this, we can again express $Y_{reh,IR}$ as a function of the freeze-out Yield using 
\begin{eqnarray}
     \left.\frac{n_{eq}^{2}\langle\sigma v\rangle a^{3}}{H}\right|_{RH} = \left.\frac{n_{eq}^{2}\langle\sigma v\rangle a^{3}}{H}\right|_{FO,h}\left(\frac{T_{RH}}{T_{\times}}\right)^{\frac{\xi(k+2)(n+6)-6(k+1)}{k+2}}\nonumber\\ \times\left(\frac{T_{\times}}{T_{FO,h}}\right)^{\frac{(k+2)(n+6)-6(k+1)}{k+2}}
\end{eqnarray}
which can again be expressed in terms of the freeze-out Yield using Eq.~\eqref{eq:thermal eq}. Gathering all the results, the production in each cases can now be summarized in the following way
\begin{widetext}
    \begin{equation}
    \label{eq:production yield}
        Y_{\chi}(a_{RH}) = Y_{FO}\times\begin{cases}
            1+\frac{3k}{\xi(n+6)(k+2)-6(k+1)}\left(\frac{T_{\times}}{T_{FO,h}}\right)^{\frac{6}{k+2} +n}\,&\mathrm{UV}\,,\\
            1 +\frac{3k}{6(k+1)-\xi(n+6)(k+2)}\left(\frac{T_{\times}}{T_{FO,h}}\right)^{\frac{6}{k+2} +n}\left(\frac{T_{RH}}{T_{\times}}\right)^{6+n - \frac{6(1+k)}{\xi(k+2)}}&\mathrm{IR}\,,m_{\chi}<T_{RH}\,,\\
            1 +\frac{3k}{6(k+1)-\xi(n+6)(k+2)}\left(\frac{T_{\times}}{T_{FO,h}}\right)^{\frac{6}{k+2} +n}\left(\frac{m_{\chi}}{T_{\times}}\right)^{6+n - \frac{6(1+k)}{\xi(k+2)}}&\mathrm{IR}\,,T_{RH}< m_{\chi}<T_{\times}\,,
        \end{cases}
    \end{equation}
\end{widetext}
The IR case with $m_{\chi}>T_{RH}$ is computed in a similar way as for the standard IR case except that the integration in Eq.~\eqref{eq:dark matter production} will go from $T_{\times}$ to $m_{\chi}$ \footnote{ Note that the result for $m_\chi > T_{RH}$ in Eq.~(\ref{eq:production yield}) accounts for the relativistic production and is consistent with previous studies of UFO and freeze-in during reheating. However, for the IR case, there is in general significant additional production that occurs in the non-relativistic regime for $T < m_\chi$. For instance, for a heavy $Z'$ portal UFO model, we found that the non-relativistic production may be $\mathcal{O}(30-200)$-fold greater than the relativistic one \cite{Henrich:2026tox}. However, this additional non-relativistic production is model-dependent, so we neglect it here. The general effect of the latter on our results can easily be visualized as a shift of each contour for a particular $\Lambda$ down in the $(m_\chi,T_{RH})$ plane.}. In the UV case we see that the second term is negligible except when $T_{\times}\sim T_{FO,h}$ ie. when the system is transitioning between GUFO to UFO. In the IR case though the second term is a bit more complex as it will be again a duration competition between the duration of the $T\sim a^{-1}$ and the $T\sim a^{-\xi}$ era. The term $T_{\times}/T_{FO,h}$ tends to reduce the contribution while the term $T_{RH}/T_{\times}$ tends to make it grow supporting our intuition. In a similar way as for the potential rethermalization, the IR case is then not guaranteed to have a significant IR production as a decoupling deep in the gravitational radiation will reduce the production term potentially leading to a UV like production regardless of $\xi$.
The limiting case where the IR production dominates can then be recast as a condition on the freeze-out temperature using Eq.~\eqref{eq:production yield}

\begin{align}
\label{eq:UVIR condition}
    \frac{T_{FO,h}}{T_{\times}}\ll\left(\frac{3k}{6(k+1)-\xi(n+6)(k+2)}\right)^{\frac{k+2}{(k+2)(n+6)-6(k+1)}}\\\times\left(\frac{T_{RH}}{T_{\times}}\right)^{\frac{\xi(k+2)(n+6)-6(k+1)}{\xi(k+2)n + 6\xi }} \nonumber\,, 
\end{align}
which can be easily adapted to the case $m_{\chi}>T_{RH}$ by the replacement $T_{RH} \rightarrow m_{\chi}$.
We display this condition assuming $m_{\chi}<T_{RH}$ on Fig.~\ref{fig:parameter_space} when relevant. The process can then be IR dominated below the dashed lines and above the full line. Note that for the case $k=2 ~{\rm and} ~ n=0$ the IR limit for GUFO lies above the range of the plot so no UV GUFO is possible in this case. We see that IR GUFO dominates for high enough reheating temperature while UV GUFO remains mostly valid for small reheating temperature and high BSM scale. This can be understood easily in the case $k=4$. There the duration for the reheating contribution to dominate is independant of the reheating temperature. Hence, increasing  $T_{RH}$ will only reduce the ratio $a_{\times}/a_{FO,h}$ transitioning to IR GUFO since the source term will redshift less between decoupling and crossing. 

 In the above derivations of the abundance $Y_\chi(a_{RH})$, we have required that $T_{FO}>T_\times$ such that GUFO occurs. For completeness, we also now report the final abundance when $T_{FO}<T_{\times}$, which corresponds to the generalized result for standard UFO for arbitrary $\xi$. In the case of standard IR UFO, the final abundance for arbitrary $\xi$ can be readily obtained from the IR terms in Eq.~(\ref{eq:production yield}), since the post-freeze-out contributions sourced by the radiation bath during the second phase of reheating ($T<T_\times$) are the same for standard IR UFO and GUFO. However, the final abundance for standard UV UFO cannot be readily obtained from Eq.~(\ref{eq:production yield}). The generalized result for the final abundance for standard UV UFO for arbitrary $\xi$ is
\begin{align}
\label{eq:YuvUFO}
Y_{\chi}(a_{RH})&\!=\!Y_{FO,reh}\!+\!\left(\frac{3k}{\xi(n\!+\!6)(k\!+\!2)\!-\!6k\!-\!6}\right)Y_{FO,reh} \nonumber \\
& =\left(\frac{\xi(n+6)(k+2)-3k-6}{\xi(n+6)(k+2)-6k-6}\right)Y_{FO,reh},
\end{align} where $Y_{FO,reh}$ is given by Eq.~(\ref{eq:YUFO}). We note that this result is consistent with the previously published result for fermionic reheating in \cite{Henrich:2025gsd}, as required. Importantly, the abundance in Eq.~(\ref{eq:YuvUFO}) is sensitive to both $\Lambda$ and $n$, which is distinct from the analogous case for UV GUFO.



\subsection{Producing the right relic abundance.}

\begin{figure*}[t]
    \centering
    \includegraphics[width=0.48\textwidth]{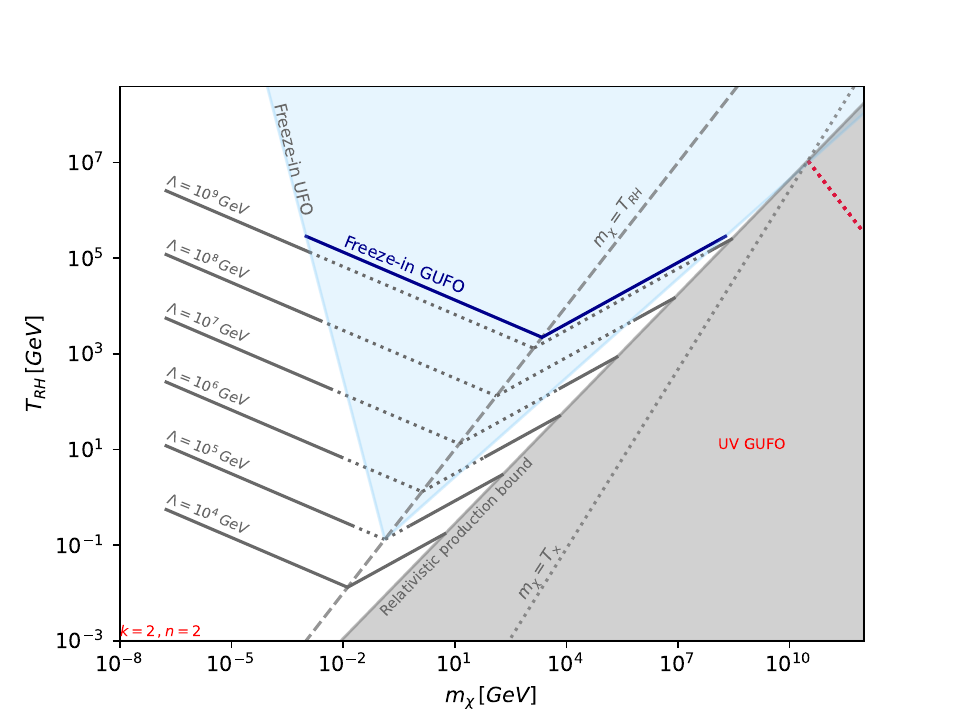}
    \hfill
    \includegraphics[width=0.48\textwidth]{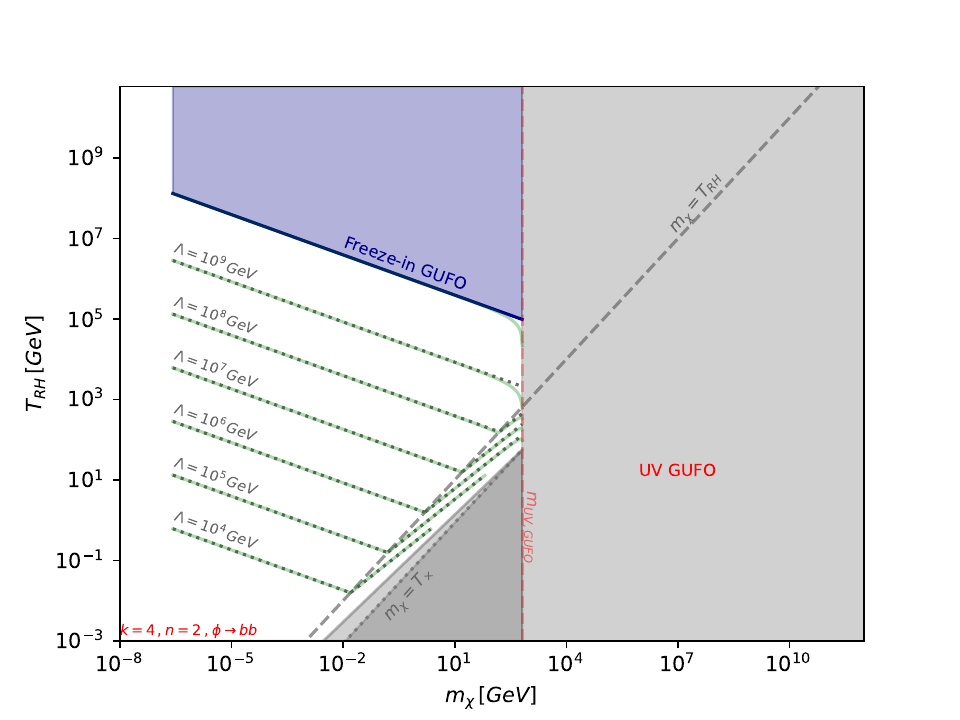}
    \caption{\justifying$m_{\chi}$ vs $T_{RH}$ plane for consistent relic abundance with $\Omega_{\chi}h^{2}$ for $k=2$, $n=2$ (left panel) and for $k=4$, $n=2$, $\phi\rightarrow bb$ (right panel). In each case we display in solid line the standard UFO prediction while the GUFO prediction is showed in dotted lines. Additionally, we present relevant limits such as $m_{\chi} = T_{RH}$, $m_{\chi}=T_{\times}$ and the standard UFO limit for the left panel to facilitate comparison with Ref.~\cite{Henrich:2025gsd}. The boundary between the dotted and solid portions of the contours can be obtained by setting $T_{FO}=T_{\times}$ and enforcing the correct relic abundance. On the right panel in green we add a numerical optimization of $\Omega_{\chi}h^{2}=0.12$ using the complete expression for the produced yield from Eq.~\eqref{eq:production yield}. Finally we display the thermalization condition requiring $T_{FO,h}<\Lambda$ in dark blue.}
    \label{fig:relic abundance}
\end{figure*}

Now that we computed the production in the different regimes we are in position to compute the dark matter relic abundance. Let us start with the simplest case of UV GUFO, namely when the potential IR contribution never dominates. This scenario can happen in 3 regimes. The two simplest ones are if standard UFO would be UV or if Eq.~\eqref{eq:UVIR condition} is not satisfied with a standard IR UFO stemming from a short enough period where $T\sim a^{-\xi}$. Otherwise, one could imagine a dark matter candidate whose mass satisfies $T_{FO,h}>m_{\chi}>T_{\times}$. In this case the dark matter will become non-relativistic prior to a potential IR production implying a Boltzmann suppression in the source term from Eq.~\eqref{eq:dark matter production}.
In all these cases, the production is independent from the exact interaction leading to a simple relic abundance of the form
\begin{equation}
\label{eq:UVGUFO}
    \Omega_{\chi,UV}h^{2}\simeq \frac{h^{2}}{\rho_{c}}\frac{g_{\chi}\zeta(3)}{\pi^{2}\mathcal{T}^{\frac{3(1-\xi)}{\xi}}}\left(\frac{g_{0}}{g_{RH}}\right)T_{0}^{3}m_{\chi}T_{RH}^{\frac{4-k}{k}}\,,
\end{equation}
where here $\rho_{c}\simeq 8.1\times10^{-47} h^{2}\,\mathrm{GeV}^{4}$ is the critical energy density, $g_{0}\simeq3.91$ and $g_{RH}\simeq 106.75$ are the number of degree of freedom today and at reheating respectively. From this relation emerges an independence on the reheating temperature for the specific case $k=4$. This feature is to be expected as in this case we have $\rho_{\phi}/\rho_{R}\sim cst$ when the gravitationally produced radiation dominates. In this case, the duration of the $T\sim a^{-\xi}$ era diluting the dark matter is independent from the reheating temperature. This can be seen in Eq.~\eqref{eq:Tcross} which reduces to a linear relation for $k=4$. In this specific case, the mass of the dark matter candidate is completely fixed as $\mathcal{T}^{\frac{1-\xi}{\xi}}$ is independent of the reheating channel. This relation gives $m_{\chi}\sim 10^{3} \,\mathrm{GeV}$ independently of $n$. This is one of our main findings as UV GUFO is a new feature arising solely from the initial radiation production through gravitational processes. 
If instead $k\neq4$ the relevant quantity is the product $m_{\chi}T_{RH}^{\frac{4-k}{k}}$ implying a higher mass for a smaller reheating temperature in the case $k=2$ but we will not develop more on this case as it is potentially available only in the case $k=2,\,n=2$ and is in the end forbidden by other constraints as it shown in Fig.~\ref{fig:relic abundance}.

For IR GUFO, we will focus on the two cases $k=2,\,n=2$ and $k=4,\,n=2,\,\phi\rightarrow bb$ as they have the wider parameter space. Using Eqs.~\eqref{eq:Tcross}\eqref{eq:genericTfoh}\eqref{eq:YGUFO} and Eq.~\eqref{eq:production yield} these cases can be evaluated to
\begin{align}
    \Omega_{\chi}^{m_{\chi}<T_{RH}}h^{2} =\frac{h^{2}}{\rho_{c}}m_{\chi}\frac{g_{0}}{g_{RH}} T_{0}^{3}\frac{2}{\sqrt{3\alpha}}\left(\frac{g_{\chi}\zeta(3)}{\pi^{2}}\right)^{2}\\\times\begin{cases}
        \frac{M_{P}T_{RH}^{3}}{\Lambda^{4}}&\mathrm{for}\,k=2,\,n=2\\
        \frac{1}{2}\frac{M_{P}T_{RH}^{3}}{\Lambda^{4}}&\mathrm{for}\,k=4,\,n=2,\,\phi\rightarrow bb
    \end{cases}\nonumber\,.
\end{align}
\begin{align}
    \Omega_{\chi}^{m_{\chi}>T_{RH}}h^{2} =\frac{h^{2}}{\rho_{c}}\frac{g_{0}}{g_{RH}} T_{0}^{3}\frac{2}{\sqrt{3\alpha}}\left(\frac{g_{\chi}\zeta(3)}{\pi^{2}}\right)^{2}\\\times\begin{cases}
        \frac{M_{P}T_{RH}^{7}}{m_{\chi}^{3}\Lambda^{4}}&\mathrm{for}\,k=2,\,n=2\\
        \frac{1}{2}\frac{M_{P}T_{RH}^{15}}{m_{\chi}^{11}\Lambda^{4}}&\mathrm{for}\,k=4,\,n=2,\,\phi\rightarrow bb
    \end{cases}\nonumber\,.
\end{align}

We display these results in Fig.~\ref{fig:relic abundance}.  The standard results from UFO are visible in full grey line while the GUFO contribution is displayed in dotted grey. 
In addition, we present the thermalization condition for GUFO to happen in our setup and the $m_{\chi}=T_{\times}$ line relevant for UV GUFO. For an easier comparison with Ref.~\cite{Henrich:2025gsd}, we display the standard UFO thermalization limit and the relativistic production bound in light blue and grey, respectively. Let us now make a set of comments starting on the left panel for the case $k=2$, $n=2$. This case is already allowed for standard UFO but is mainly constrained by the thermalization limit represented in light blue, corresponding to the condition $T_{FO,reh}<\Lambda$. There, using Eq.~\eqref{eq:genericTforeh} the dependence on $T_{RH}$ gives a rapid decrease of the thermalization condition. In this case, the main effect of GUFO is to open the parameter space, once $\Lambda \lesssim 10^{9}\,\mathrm{GeV}$, to higher masses in the $m_\chi <T_{RH}$ region and lower masses in the $m_\chi > T_{RH}$ region. This is easily understood from Eq.~\eqref{eq:genericTfoh}. As for GUFO, the freeze-out temperature does not depend on the reheating temperature, requiring $T_{FO,h}=\Lambda$ for thermalization only fixes the BSM scale to a specific value, therefore transforming the limit into a contour line with $\Omega_{\chi}h^{2}=0.12$. We display in this case the UV GUFO production in dotted red valid only for $m_{\chi}>T_{\times}$ in this case.  However, the UV GUFO regime is only reached for dark matter masses which exceed the limit on relativistic DM production sourced by the thermal bath. This limit exists simply due to the fact that the DM number density cannot exceed the equilibrium number density in the relativistic regime ($T\gtrsim m_\chi$) such that we require $n_\chi(a_m)\leq n_{\rm \chi,eq}(a_m)$ ($a_{m}$ being the scale factor at $T=m_{\chi}$). Recall that our analytic estimates of the abundance above involved a lower limit of integration on the temperature, $T=m_\chi$, when $m_\chi>T_{\rm RH}$, such that the DM number density at $a_m$ simply redshifts proportional to $a^{-3}$ between $a_m$ to $a_{RH}$\footnote{  When a specific model is chosen, the true cosmological upper limit on DM production sourced by the radiation bath is related to the maximum comoving equilibrium number density, which occurs at $T\simeq \frac{2}{13}m_\chi$ for $k=2$, which is in the non-relativistic regime. Because the shape of the contours in the $(T_{\rm RH},m_\chi)$ plane for fixed $\Lambda$ are model-dependent in the non-relativistic regime, we provide instead the upper limit on relativistic production.}. From this, an upper bound on the DM number density at $a=a_{RH}$ can be obtained as follows
\beq
n_\chi(a_{RH})\!=\!n_{\chi,eq}(a_{m})\left(\frac{a_{m}}{a_{RH}}\right)^3\!\simeq\! \frac{g_\chi \zeta(3)}{\pi^2}m_\chi^3\left(\frac{T_{RH}}{m_\chi}\right)^{3/\xi},
\eeq where we have written the dilution factor $\left(\frac{a_{m}}{a_{RH}}\right)^3$ as a ratio of $m_\chi$ and $T_{RH}$, where the evolution of the thermal bath is parametrized by $\xi$, and we have assumed $m_\chi < T_\times$ for the expression on the RHS. We can then enforce the correct relic abundance  to obtain expressions for this relativistic production limit in the $(m_\chi,T_{RH})$ plane. For $k=2$ and fermionic reheating, we find 
\begin{equation}
T_{RH}\!=\!\left(\frac{m_\chi^4}{\!5.88\!\times\!10^6 \!\text{ GeV}\! ^{-1}}\!\right)^{1/5}\!\left(\frac{\pi^2 (0.12)}{g_\chi \zeta(3)}\frac{427}{4 g_*(T_{RH})}\right)^{1/5}.
\end{equation} This limit is depicted in the left panel of Fig.~\ref{fig:relic abundance}.

We next examine the right panel corresponding to the case, $k=4$, $n=2$, $\phi\rightarrow bb$. This case is normally not allowed in standard UFO due to the too slow redshift of the temperature which is fortunately what makes GUFO possible in this case. This case is similar in terms of constraints to the previous one as it is again an IR production except for one major change. In particular, here we need to stop the production to the mass corresponding to UV GUFO regardless of whether $T_{RH}$ is bigger or smaller than $m_{\chi}$. To understand this effect let us take a fixed value of $\Lambda$ therefore fixing the freeze-out temperature and take a specific ordering of the mass and the reheating temperature. If $m_{\chi}<T_{RH}$ the Yield is independent of the mass, as $\Omega_{\chi}h^{2}\sim m_{\chi}Y_{\chi}$ reaching $0.12$ for high mass then requires a reduction of the Yield corresponding to a lower reheating temperature. If this temperature becomes too low, the condition from Eq.~\eqref{eq:UVIR condition} will not be met anymore and the production will shift to UV GUFO fixing the mass as previously mentioned. On the other hand if $m_{\chi}>T_{RH}$, raising the mass will tend to decrease the IR GUFO production until it no longer dominates compared to the UV one.  This feature is then generic for all cases as long as $k=4$ due to this peculiar behavior of Eq.~\eqref{eq:UVGUFO} and the dark matter mass cannot exceed $m_{\chi}\sim 10^{3}\mathrm{GeV}$.
The relativistic production bound for $k=4$ and $\phi\rightarrow bb$ is depicted in the right panel of Fig.~\ref{fig:relic abundance} and is given by
\beq
T_{RH}\!=\!\left(\frac{m_\chi^8}{5.88 \! \times \! 10^6 \!\text{ GeV}^{-1}} \!\right)^{1/9}\!\left(\frac{\pi^2 (0.12)}{g_\chi \zeta(3)}\frac{427}{4 g_*(T_{RH})}\right)^{1/9}.
\eeq

\section{Compatibility with $\Lambda\mathrm{CDM}$}
\label{sec:compatibilite LCDM}

As for all the warm dark matter models, after production, the expansion should allow for a sufficient cooling to be compatible with two extra constraints. Firstly, the dark matter should satisfy the bound on the effective number of relativistic degrees of freedom at BBN time ($T\sim 1\,\mathrm{MeV}$). Secondly, the dark matter should be compatible with the Lyman-$\alpha$ forest \cite{Irsic:2017ixq} requiring dark matter to be sufficiently slow at the time of structure formation ($T\sim 1\,\mathrm{eV}$) to generate the right size for the cosmological structures. Of course the first condition only applies to particles with mass below $1\,\mathrm{MeV}$ while the second one applies regardless of the mass. Contrary to standard UFO whose freeze-out temperature is related to the reheating temperature through Eq.~\eqref{eq:genericTforeh}, GUFO is independent of this and hence it is not guaranteed a priori that it will satisfy the above constraints which we should derive in our specific case. Both of them arise from the same relation coming from the dark sector temperature $\theta$ evaluated at a different scales after reheating. In our context it is expressed as
\begin{equation}
    \theta = 
        T_{FO,h}\frac{a_{FO,h}}{a_{\times}}\frac{a_{\times}}{a_{RH}}\frac{a_{RH}}{a}\,,\hspace{0.5cm}\mathrm{UV}\,,
\end{equation}
where here we already divided the relevant redshifts and we focus on the UV regime since the IR GUFO is exactly the same as standard UFO and therefore compatible with these constraints. Taking the variation of degrees of freedom into account the temperature-to-redshift relation will follow
\begin{equation}
    T(a) = \begin{cases}
        T_{\times}\left(\frac{g_{\times,\rho}}{g_{\rho}(T)}\right)^{\frac{1}{4}}\left(\frac{a_{\times}}{a}\right) &\mathrm{for}\,a_{max,h}\leq a\leq a_{\times}\,,\\
        T_{RH}\left(\frac{g_{RH,\rho}}{g_{\rho}(T)}\right)^{\frac{1}{4}}\left(\frac{a_{RH}}{a}\right)^{\xi}&\mathrm{for}\,a_{\times}\leq a\leq a_{RH}\,,\\
        T_{RH}\left(\frac{g_{RH,s}}{g_{s}(T)}\right)^{\frac{1}{3}}\left(\frac{a_{RH}}{a}\right)&\mathrm{for}\,a_{RH}\leq a\leq a_{0}\,,
    \end{cases}
\end{equation}
with the index $\{\rho ~ {\rm or} ~ s \}$ specifying the energy density or entropic degrees of freedom.  The dark sector temperature can then be evaluated after reheating in both cases
\begin{equation}
\label{eq:dark sector temperature}
    \theta =
        T\left(\frac{T_{RH}}{T_{\times}}\right)^{\frac{1-\xi}{\xi}}\left(\frac{g_{s}(T)}{g_{RH,s}}\right)^{\frac{1}{3}}\left(\frac{g_{\times,\rho}}{g_{FO,h,\rho}}\right)^{\frac{1}{4}} \left(\frac{g_{RH,\rho}}{g_{\times,\rho}}\right)^{\frac{1}{4\xi}}\,,
\end{equation}
which again does not depend on the BSM scale due to the intial $T\sim a^{-1}$ scaling between freeze-out and crossing. According to the previous discussion, the first constraint then requires one to evaluate Eq.~\eqref{eq:dark sector temperature} at BBN time under the bound on the extra effective number of relativistic particles $\Delta N_{eff}<0.18$ at $95\%$ CL \cite{Yeh:2022heq} suggesting
\begin{equation}
    \frac{4}{7}\left(\frac{g_{BBN}}{g_{RH}}\right)^{\frac{4}{3}}\left(\frac{g_{\times}}{g_{FO,h}}\right)\left(\frac{g_{RH}}{g_{\times}}\right)^{\frac{1}{\xi}}\frac{1}{\mathcal{T}^{\frac{4(1-\xi)}{\xi}}}<0.18T_{RH}^{\frac{4(k-4)}{3k}}\,.
\end{equation}

This condition again does not depend on the reheating temperature for $k=4$. Numerically we can check that this condition is automatically satisfied in our case if one takes $g_{RH}=g_{\times}=g_{FO,h}=106.75$ and $g_{BBN} = 10.75$. Finally, regarding the structure formation constraint, we require that the dark matter velocity is $v_{\chi}\simeq p_{\chi}/m_{\chi}\simeq \theta/m_{\chi} <2\times10^{-4}$ implying in the UV case a similar limit as in standard UFO
\begin{equation}
    m_{\chi}>5\,\mathrm{keV}\left(\frac{T_{RH}}{T_{\times}}\right)^{\frac{1-\xi}{\xi}}\,.
\end{equation}

For GUFO, the temperature ratio is solely fixed by the crossing temperature. Similarly to the standard UFO case, UV GUFO is then compatible with the structure formation constraint as this scenario requires $m_{\chi}\sim 10^{3}\,\mathrm{GeV}$. In conclusion, we see that a UV GUFO scenario will naturally be compatible with the $\Lambda$CDM cosmological model.

\section{conclusion}

The current status of dark matter is quite uncertain. The increasing number of constraints for the WIMP scenario \cite{LZ:2022lsv,XENON:2024znc,XENON:2023cxc,DARWIN:2016hyl,McDaniel:2023bju,Gaskins:2016cha} along with the lack of potential evidence for FIMP-like dark matter  motivates a revisit of the dark matter production scheme. The recent proposal of UFO \cite{Henrich:2025gsd,Henrich:2025pca,Henrich:2025sli} directly aims to find alternatives reachable by ongoing and future experiments \cite{Henrich:2026tox,Chakraborty:2026chp,Yang:2026syo}. Although relativistic freeze-out from the Standard Model bath was long believed to be incompatible with current cosmological constraints, this mechanism becomes viable in non standard cosmologies such as during reheating.
The UFO process is unique since its freeze-out temperature is unrelated to the dark matter mass, in contrast with the WIMP, while it is initially in thermal equilibrium in contrast with the FIMP scenario opening a rich phenomenology depending on the early-universe model without any dependence on a possible initial production during inflation \cite{Kolb:2023ydq}. The price to pay is then a constraining thermalization condition and the difficulty to realize this scenario with renormalizable interactions ($n=-2$) due to the limitation $n>-\frac{6}{k+2}$ for $T\sim a^{-1}$. In this report, we outlined that due to its intrinsic properties, UFO should be considered coherently with a complete model of thermal evolution. Indeed the possibility for UFO to achieve freeze-in production subsequent to its decoupling is intrinsically related to the temperature evolution as well as to the relevant interactions. Taking as an example a reheating process driven by one inflaton coupling (inflaton decays or scattering to Standard Model particles) along with the unavoidable radiation produced through gravitational portals \cite{Clery:2021bwz,Clery:2022wib}, we demonstrated that the naive approach on the decoupling condition with a specific temperature scaling is not enough to describe the UFO parameter space and that a sufficiently low reheating temperature can imply a decoupling from the gravitationally produced radiation. We found that even if the process can be IR dominated, the production can be UV like for a sufficiently high dark matter mass ($m_{\chi}>T_{\times}$) or if the decoupling happens too early compared to the crossing time. Finally, focusing on the cases $k=2$, $n=2$ and $k=4$, $n=2$ for $\phi\rightarrow bb$, we showed that for the first case GUFO extended the accessible parameter space by softening the thermalization constraint without changing the production mechanism due to the IR nature of the production. We should stress that this point is of particular interest as in our framework any model satisfying $\Lambda\lesssim10^{9}\,\mathrm{GeV}$ should then thermalize during reheating while having a freeze-in like production. This feature provides a solution to the significant overproduction problem faced by freeze-in models in high-scale inflation due to the minimal gravitational production \cite{Clery:2021bwz} without resorting to flat temperature profiles as for the Freeze-in at stronger coupling models \cite{Cosme:2023xpa,Lebedev:2024mbj,Arcadi:2024obp,Arcadi:2024wwg,Koivunen:2024vhr,Cosme:2024ndc,Feiteira:2026qme,Bertou:2026osq}. Relatedly, the many studies which investigate freeze-in dark matter in the setting of low reheating temperatures, but which do not use a full Boltzmann treatment of DM evolution during the non-instantaneous reheating period may have large regions of their parameter space which are mislabeled as freeze-in and should be re-classified as UFO/GUFO, depending on the precise DM model and reheating treatment. For instance, for Higgs portal models with fermionic DM \cite{Arcadi:2024wwg} or vector DM \cite{Khan:2025keb}, as well as the spin-2 mediated DM model in \cite{Lee:2024wes}, and the charged parent model of \cite{Goudelis:2026lyy}, the authors assume that the DM production is Boltzmann suppressed and begins at $T=T_{\rm RH}$. This treatment (and the classification of the process as freeze-in) is valid only for a restricted class of reheating models, for instance a reheating era characterized by a flat temperature profile. However, for a typical reheating era generated by inflaton decays or scattering, a full Boltzmann evolution during the non-instantaneous reheating period will often reveal that the DM would thermalize at high temperatures $T>T_{\rm RH}$ and should be classified as UFO (or in some cases non-relativistic freeze-out) in much of the relevant parameter space. As a general rule, any DM production mechanism which proceeds via a non-renormalizable operator should be checked for thermalization during the reheating era rather than using an instantaneous reheating approximation.  Moreover, we found that even if UFO was ordinarily forbidden for the standard $\phi\rightarrow bb$ temperature evolution, GUFO makes this reheating channel accessible, with a parameter space as large as that of standard UFO for $\phi\rightarrow f\bar{f}$. This is of particular interest as the bosonic decay channel is the least affected by fragmentation, contrary to $\phi\rightarrow f\bar{f}$, which is the most affected one \cite{Garcia:2023dyf}. Additionally, in the case of Higgs final state, it can solve the stability of the potential if $T_{RH}\gtrsim 10^{6}\,\mathrm{GeV}$ and easily accommodate baryogenesis \cite{Cado:2025orb,Cado:2022evn} with the adjunction of a bare mass term irrelevant for our analysis. More broadly, as long as $k=4$, the dark matter mass cannot exceed $m_{\chi}\sim 10^{3}\,\mathrm{GeV}$ and be consistent with UFO regardless of the UV or IR nature of the production. \\
Let us close this section by noting that, in addition to the gravitationally produced radiation after the end of inflation, one should also take into account another model-independent source of radiation originating from the Higgs field. The latter, either as spectator or even non-minimally coupled, admits a condensation \cite{Kunimitsu:2012xx} which could have similar energy density with the gravitational thermal bath. Then, according to its oscillation/decay time \cite{Figueroa:2016ojl}, it may drastically alter $a_\times$, which in turn may enable UFO to occur for an even wider range of parameters. However, since such a case deserves its own analysis, we leave it for a continuation of this work.

\section*{ACKNOWLEDGEMENTS}

The authors would like to thank Yann Mambrini for useful discussions and comments on the first version of the manuscript. M.G. and S.E.H. acknowledge support by Institut Pascal at Université Paris-Saclay during the Paris-Saclay
Astroparticle Symposium 2025.
The work of F.K. is supported in part by the National Natural Science Foundation of China under grant No. 12342502.

\appendix
\section{Inflaton dynamics and decay widths}
\label{app:inflaton}
The purpose of this appendix is to give an overview of the necessary reheating considerations regarding inflaton oscillations. We refer the reader to Ref.~\cite{Garcia:2020wiy} for a complete treatement of reheating. In an expanding background, the homogeneous inflaton $\phi$ obeys the Klein Gordon equation 
\begin{equation}
\label{Eq:eom_inflaton}
\ddot{\phi}+3H\dot{\phi} +\frac{dV_\phi}{d\phi} \approx 0\,.
\end{equation}
The solution to Eq.~\eqref{Eq:eom_inflaton} can then be parametrized as follows for a potential given by Eq.~\eqref{eq:inflaton potential}
\begin{equation}
\phi(t) = \phi_0(t)\underbrace{\sum_{\nu} \mathcal{P}_{\nu}(t)e^{-i\nu \omega t}}_{\mathcal{P}}\ ,
\label{Eq:generalphi}
\end{equation}
where $\phi_0$ is the envelope of the field and the $\mathcal{P}_{\nu}$ are the Fourier coefficients of the pseudo-harmonic function $\mathcal{P}$. Since the time scale for oscillations  $\mathcal{O}(m_{\phi})$, with
\beq
m_{\phi}^2=\left.\frac{\partial^2 V_\phi}{\partial \phi^2}\right|_{\phi=\phi_0}
=k(k-1)\lambda M_P^2\left(\frac{\phi_0}{M_P}\right)^{k-2}\,,
\eeq
is fast compared to the expansion rate of the universe, $H$, we can decompose the equation of motion (\ref{Eq:eom_inflaton}) between the envelope (the amplitude of oscillations) and the pseudo-periodic part
\begin{align}
    \dot{\phi_0} &= -\frac{6H}{k+2}\phi_0, \label{eq:envelope}\\
    \dot{\mathcal{P}}^{2} &= \frac{2m_{\phi}^{2}}{k(k-1)}(1-\mathcal{P}_{}^{k}). \label{eq:harmonics}
\end{align}

Solving Eqs.~\eqref{eq:envelope} and \eqref{eq:harmonics} leads to a decaying envelope and an oscillating frequency of the form
\begin{align}
    \phi_0 &= \phi_{\rm end} \left(\frac{a_{\rm end}}{a}\right)^{\frac{6}{k+2}},
    \label{phi0}\\
    \omega &\simeq m_{\phi} \ \sqrt{ \frac{ \pi k }{ 2(k - 1) }}\frac{ \Gamma\left( \frac{1}{2} + \frac{1}{k} \right) }{ \Gamma\left( \frac{1}{k} \right) } \,,
\end{align}
Where the various inflaton modes $\mathcal{P}_{\nu}$ yield an energy $E_{\nu} = \nu\omega$ sourcing the radiation production described in Eq.~\eqref{eq:gravitational rad}. 

Properly achieving reheating requires additional couplings, in order to source the radiation, on top of the free field behavior we just exposed. Here we consider couplings of the form of Eq.~\eqref{eq:inflaton channels}. The corresponding inflaton decay rates can then be computed in the form \cite{Garcia:2020wiy}
\begin{equation}
\label{eq:inflaton decay rates}
    \Gamma_{\phi}=
    \begin{cases}
        \frac{y_{eff}^{2}}{8\pi}m_{\phi}      & \phi \rightarrow \bar{f} f\\
        \frac{\mu_{eff}^{2}}{8\pi m_{\phi}}      & \phi \rightarrow bb\\        \frac{\sigma_{eff}^{2}\rho_{\phi}}{8\pi m_{\phi}^{3}}      & \phi\phi \rightarrow bb\,.
    \end{cases}
\end{equation}

    The mapping to the generic form in Eq.~\eqref{eq:generic gammaphi} gives
    \begin{equation}
\label{eq:gammaphi}
    \gamma_\phi =
\begin{cases}
\dfrac{\sqrt{k(k-1)}\,\lambda^{1/k} M_P}{8\pi}\, y_{\mathrm{eff}}^{2},
& \phi \rightarrow \bar{f} f, \\[1.2ex]
\dfrac{\mu_{\mathrm{eff}}^{2}}
{8\pi \sqrt{k(k-1)}\,\lambda^{1/k} M_P},
& \phi \rightarrow b b, \\[1.2ex]
\dfrac{\sigma_{\mathrm{eff}}^{2} M_P}
{8\pi \,[k(k-1)]^{3/2}\,\lambda^{3/k}},
& \phi\phi \rightarrow b b ,
\end{cases}
\end{equation}
with the $l$ values respectively given by
\begin{equation}
    \ell =
\begin{cases}
\dfrac{1}{2} - \dfrac{1}{k},
& \phi \rightarrow \bar{f} f, \\[1ex]
\dfrac{1}{k} - \dfrac{1}{2},
& \phi \rightarrow b b, \\[1ex]
\dfrac{3}{k} - \dfrac{1}{2},
& \phi\phi \rightarrow b b 
\end{cases}
\end{equation}
where the effective couplings $y_{eff},\,\mu_{eff}$ and $\sigma_{eff}$ are different from the lagrangian couplings as they include kinematic effects that we will neglect for simplicity. From there, the reheating dynamics can be solved as presented in Sec.~\ref{Sec:reheating} leading to Eq.~\eqref{eq:reheating profile}.
\clearpage
\bibliography{biblio}
\end{document}